%% file: sept18_halobiaspaper.tex
\documentclass[a4paper,11pt]{article}
\pdfoutput=1 
\usepackage{jcappub} 

\usepackage{multirow}
\usepackage[T1]{fontenc} 
\usepackage{graphicx}
\usepackage{todonotes}
\usepackage{caption}
\usepackage{subcaption}

\def\be{\begin{equation}}
\def\ee{\end{equation}}
\def\bi{\begin{itemize}\renewcommand{\labelitemi}{$\circ$}}
\def\ei{\end{itemize}}
\def\ben{\begin{enumerate}}
\def\een{\end{enumerate}}
\def\bt{\begin{tabular}}
\def\et{\end{tabular}}
\def\bc{\begin{center}}
\def\ec{\end{center}}
\def\la{\label}
\def\bh{b_{halo}}

\def\bea{\begin{eqnarray}}
\def\eea{\end{eqnarray}}
\def\la{\langle}
\def\ra{\rangle}

\begin{document} 
\title{Non-Gaussian Shape Discrimination with Spectroscopic Galaxy Surveys}

\author{Joyce Byun and Rachel Bean}
\affiliation{Department of Astronomy, \\Cornell University, \\Ithaca, NY 14853, USA}

\emailAdd{byun@astro.cornell.edu}
\emailAdd{rbean@astro.cornell.edu}

\abstract{
We consider how galaxy clustering data, from Mpc to Gpc scales,  from upcoming large scale structure surveys, such as Euclid and DESI, can provide discriminating information about the bispectrum shape arising from a variety of inflationary scenarios. Through exploring in detail the weighting of shape properties in the calculation of the halo bias and halo mass function we show how they probe a broad range of configurations, beyond those in the squeezed limit, that can help distinguish between shapes with similar large scale bias behaviors.

We assess the impact, on constraints for a diverse set of non-Gaussian shapes, of galaxy clustering information in the mildly non-linear regime, and surveys that span multiple redshifts and employ different galactic tracers of the dark matter distribution. Fisher forecasts are presented for a Euclid-like spectroscopic survey of H$\alpha$-selected emission line galaxies (ELGs),  and a DESI-like survey, of luminous red galaxies (LRGs) and [O-II] doublet-selected ELGs, in combination with Planck-like CMB temperature and polarization data. 

While ELG samples provide better probes of shapes that are divergent in the squeezed limit, LRG constraints, centered below $z<1$, yield stronger constraints on shapes with scale-independent large-scale halo biases, such as the equilateral template. The ELG and LRG samples provide complementary degeneracy directions for distinguishing between different shapes. For H$\alpha$-selected galaxies, we note that recent revisions of the expected H$\alpha$ luminosity function reduce the halo bias constraints on the local shape, relative to the CMB. For galaxy clustering constraints to be comparable to those from the CMB, additional information about the Gaussian galaxy bias is needed, such as can be determined from the galaxy clustering bispectrum or probing the halo power spectrum directly through weak lensing. 
If the Gaussian galaxy bias is constrained to better than a percent level then the LSS and CMB data could provide complementary constraints that will enable differentiation of bispectrum with distinct theoretical origins  but with similar large scale, squeezed-limit properties.
}

\maketitle
\flushbottom

\input{sept18_halobiaspaper_intro}
\input{sept18_halobiaspaper_formalism}

\input{sept18_halobiaspaper_results}

\input{sept18_halobiaspaper_discussion}

\acknowledgments
The authors would like to thank  Nishant Agarwal, Tommaso Giannantonio, Shirley Ho, and Sarah Shandera for useful inputs and discussions. The work of JB and RB is supported by NASA ATP grants NNX11AI95G and NNX14AH53G, NASA ROSES grant 12-EUCLID12- 0004, NSF CAREER grant 0844825 and DoE grant DE-SC0011838.

\bibliographystyle{JHEP}
\bibliography{allrefs}

\end{document}

%% file: sept18_halobiaspaper_intro.tex
\section{Introduction}
\label{sec:intro}

The physical mechanism behind inflation still remains an open question, despite the successes of the inflationary paradigm.  An abundance of physically-motivated theories of inflation exist that employ very different physics to drive inflation, ranging from the simplest models with a single slowly-rolling inflaton field, to those with multiple fields, non-canonical kinetic terms, or originating from excited initial states, to broadly name a few. In general, however, each model predicts a distinct `thumbprint' signature in the statistics of the generated primordial perturbations due to the particular self-interactions of the inflationary field(s) that give rise to a scale-dependence in deviations from Gaussianity in tracers of the dark matter density and Cosmic Microwave Background (CMB) temperature and polarization perturbations.

Recent results from the Planck survey of the CMB have put the strongest constraints yet on the potentially non-Gaussian statistics of primordial perturbations \cite{Ade:2013uln,Ade:2013ydc} from the imprint on the observed multipole bispectrum. The primordial signal appears as changes in the dark matter halo bias and halo mass function at later cosmological epochs.  By exploiting how galaxies are distributed with respect to these halos, we can extract shape constraints from galaxy surveys. Upcoming experiments, such as the Large Synoptic Survey Telescope (LSST) \cite{Abell:2009aa}, Euclid \cite{Refregier:2010ss}, and DESI \cite{Levi:2013gra}, will capture the statistics of the evolved late-time perturbations through the distribution of large-scale structure (LSS). These probe non-Gaussianity, potentially even more strongly than Planck (e.g. \cite{Carbone:2008iz,Fedeli:2010ud,Amendola:2012ys}). Furthermore, the CMB and LSS observables probe the primordial fluctuations differently, offering the enticing possibility of using them together as complementary data sets to strengthen existing constraints (e.g. \cite{Verde:2010wp}).

The simplest way for a distribution to be non-Gaussian is to have a non-zero 3-point correlator, i.e. bispectrum or `shape'.  Typically the amplitude of primordial $N$-point correlators  grows smaller as $N$ increases, so that the bispectrum constraints are the principal N-point non-Gaussian correlation statistic investigated.  An exciting realization in inflationary theory is that discerning the scale-dependence, or `shape', of the bispectrum could provide a direct insight into the inflationary mechanism, through how non-Gaussianity is generated \citep{Fergusson:2009nv,Liguori:2010hx}.  Taking a specific bispectrum template and comparing it to data is a sensible approach to assess if the theory is consistent, or if it possesses new properties that might be observable with appropriately designed future surveys. Three templates have been studied extensively: the local shape that arises in multi-field inflation \citep{Gangui:1993tt,Verde:1999ij,Komatsu:2001rj}, and the equilateral \citep{Creminelli:2005hu} and orthogonal shapes \citep{Senatore:2009gt} that derive from non-trivial kinetic terms in the inflationary action.

Recent theoretical developments have shown there are a far wider population of bispectrum shapes, including those from fast-roll inflation \citep{Chen:2006nt,Khoury:2008wj,Noller:2011hd,Ribeiro:2012ar,Noller:2012ed}, quasi-single field inflation \citep{Chen:2009we,Chen:2009zp}, warm inflation \citep{Gupta:2002kn,Moss:2007cv}, and non-Bunch-Davies or excited initial states \citep{Chen:2006nt,Holman:2007na,Meerburg:2009ys,Agarwal:2012mq}.  There are also hybrids of multi-field and non-slow-roll models \citep{Langlois:2008qf,Arroja:2008yy,RenauxPetel:2009sj}, and the inclusion of isocurvature modes in the non-Gaussian correlations \citep{Langlois:2011zz,Langlois:2011hn,Langlois:2012tm}. The large number of alternative theories motivates looking for distinctive characteristics with which to classify bispectra and reconstruct what we can know about the underlying theory in an unbiased way from the data.  One such characteristic in particular has gained prominence in the literature: while the bispectra can have very different shapes, meaning their signal is weighted towards different configurations of the three wavenumbers in (Fourier) $k$-space, their properties in the `squeezed limit' when one of the three length scales contributing to the 3-point function becomes much larger than the other two, can signal whether inflation is derived from a single-field or multi-field model or from a non-Bunch-Davies vacuum.

For a preferred model with a particular bispectrum $B_\Phi(k_1,k_2,k_3)$, one can forecast or measure a constraint on the amplitude $f_{NL}$ of the model's shape at a particular pivot value of $(k_1,k_2,k_3)$.  Then, under the assumption that this is the right model, we can extend the fixed shape to infer the amplitude of the bispectrum at all other $k$-configurations.  Constraints obtained in this way thus assume a strong theoretical prior, as they do not account for our uncertainty in knowing which model is the right one. In contrast, an alternative and more conservative way of getting constraints on inflation is to assume less theoretical bias, and attempt to use the data itself to reconstruct the general primordial shape and its potentially complicated $k$-dependence. If there is a detection of a non-zero $f_{NL}$ given one template, e.g. if $f_{NL}^{local}\neq 0$, this could  either indicate evidence of local non-Gaussianity as might arise in multifield inflation, or that another inflationary model has generated non-Gaussianity whose signal leaks into, or is partially described by, the local template. These dual questions of distinguishing between model shapes and model-independent constraints on shapes have been considered recently in the context of CMB bispectrum data by WMAP9  \cite{Bennett:2012fp} and Planck \cite{Ade:2013ydc} .

In this paper we explore the ability of upcoming galaxy surveys and their measured galaxy power spectra, to obtain general shape constraints and distinguish between shapes, alone and in combination with CMB data. To do this, we consider the effects that different shapes have on dark matter halo statistics through the halo bias and halo mass function, with a particular focus on which $k$-configurations of the primordial shape are primarily probed by the halo statistics. By implementing the halo model, we  forecast Euclid- and DESI-like constraints on $f_{NL}$ for a variety of specific templates, as well as consider the distinguishability and degeneracies between combinations of these shapes. This work builds on prior work considering general shape constraints arising from CMB data \cite{Bennett:2012fp,Byun:2013jba,Ade:2013ydc}, and through using basis reconstruction to study general non-Gaussian shapes, provides an alternative approach to other studies of non-Gaussian halo bias constraints in the literature, that used scale dependent modifications to the local template \cite{Becker:2010hx,Becker:2012yr} and theoretically motivated two-parameter models \cite{Norena:2012yi,Sefusatti:2012ye}, to constrain the squeezed limit properties of bispectra.

The paper is organized as follows:  Section \ref{sec:formalism} describes the formalism behind our analysis, including an outline of the basis function approach used to describe general non-Gaussian  shapes,  the halo bias and halo mass function modeling, and the Fisher analysis. Our findings, centered on the potential to use LSS data to constrain shapes beyond the squeezed-limit, are presented in Section \ref{sec:results} and in Section \ref{sec:discussion} we summarize our conclusions and their implications.

%% file: sept18_halobiaspaper_formalism.tex
\section{Formalism}
\label{sec:formalism}

This section outlines the main theoretical assumptions used in the analysis. Section \ref{subsec:templates} outlines the approach we use to describe general bispectrum shapes, and   \ref{subsec:ng_halos} describes how the halo bias and halo mass function are altered in the presence of primordial non-Gaussianity.
Sections \ref{subsec:halomodel} and  \ref{subsec:galmodel} respectively summarize how halo statistics are translated into halo
and galaxy power spectra within the framework of the halo model.
Finally,  \ref{subsec:fisher_approach} outlines the Fisher matrix formalism we use to forecast constraints for upcoming galaxy surveys, such as Euclid and DESI.

\subsection{Bispectrum shape families and templates}
\label{subsec:templates}

Theoretical predictions for the primordial fluctuations created by different inflationary models are typically given by $N$-point correlators, such as bispectra, and their specific $k$-dependence.
A primordial bispectrum, $B_\zeta$, is defined through the 3-point correlation function of (comoving) curvature perturbations in Fourier space,
\be
\la \zeta(\vec{k}_1)\zeta(\vec{k}_2)\zeta(\vec{k}_3) \ra 
= (2\pi)^3 \delta(\vec{k}_1+\vec{k}_2+\vec{k}_3) B_\zeta(k_1,k_2,k_3).
\ee
Each model's bispectrum can then be parametrized in terms of an amplitude, $f_{NL}$, and shape, $S_\zeta(k_1,k_2,k_3)$, that are related by a normalization factor, $N$, chosen such that the shape is equal to unity at a pivot scale $\vec{k}_0$,
\be
(k_1 k_2 k_3)^2 B_\zeta(k_1,k_2,k_3) = N f_{NL} S_\zeta(k_1,k_2,k_3),
\ee
where $N \equiv 72 \pi^4 \Delta_\zeta^4(k_0)/5$, and $\Delta_\zeta^2(k)$ is the dimensionless primordial power spectrum.

Each model's shape is unique, but many shapes exhibit similar features, motivating us to roughly categorize shapes into families of shapes that peak when $k_1$, $k_2$, and $k_3$ form squeezed, equilateral, flattened, etc. triangles.
Thus some of the most widely studied shapes are templates that approximate the shapes within each family, e.g. the local, equilateral, orthogonal, and enfolded templates,
where the degree of similarity between shapes, or between true shapes and templates, can be quantified by a inner product statistic \cite{Fergusson:2009nv}.
These templates are meant to be used as indicators of more complex shapes that arise from models with, for example, multiple fields (local template), higher derivative terms (equilateral and orthogonal templates), or non-trivial vacuum states (enfolded template).

Going beyond shapes characterized by a single template that is chosen to reflect a preferred model, a model-independent way to parametrize a general shape is to choose a basis of shape functions and assume that the underlying true shape is well-approximated by a linear combination of the basis, such as in \cite{Fergusson:2009nv,Byun:2013jba}.
We utilize the separable $\{\mathcal{K}_n\}$ basis set introduced in \cite{Byun:2013jba}.
Each basis function is defined as 
\bea
{\cal K}_{n}(k_1,k_2,k_3) &\equiv& \frac{1}{{\cal N}_{n} k_0^{2(n_s-1)}}\left[k_1^{p'} k_2^{r'} k_3^{s'}+ \{prs\} \operatorname{ perms}\right],
\eea
where ${\cal N}_{n}$ is the number of distinct permutations of $\{p,r,s\}$, and $p', q'$ and $r'$ are defined as
\bea
p' &\equiv& 2+\frac{(p-2)(4-n_s) }{ 3}.
\eea  
To create near scale-invariance, the powers must satisfy $p+r+s=0$.

Bispectrum separability is not strictly necessary to efficiently compute the resulting halo bias and halo mass function corrections using \eqref{eq:bsd}, \eqref{eq:bsi}, and \eqref{eq:Rng} as we describe below. It is, however, a useful approach if one wants to consistently consider constraints on general shapes, and mesh together the squeezed limit and broader shape properties, from LSS and the CMB analyses, as separability is essential for the latter.
Separability also allows for computationally efficient simulations that verify our theoretical understanding of how different LSS probes depend on non-Gaussianity, e.g. for nonlinear power spectra and halo mass functions \cite{Wagner:2010me}, halo bias \cite{Wagner:2011wx}, and matter bispectra and higher-order correlators \cite{Fergusson:2010ia}.

The $\{\mathcal{K}_n\}$ basis can be used to describe many nearly scale-invariant shapes in the literature precisely using only a relatively small set of 7 basis functions,  
\be
(k_1 k_2 k_3)^2 B_\zeta(k_1,k_2,k_3)  
= N S_\zeta(k_1,k_2,k_3) \approx N\sum_{n=0}^{6} \alpha_n{\cal K}_n(k_1,k_2,k_3).
\ee
The local, equilateral, orthogonal, and enfolded templates can be expressed exactly in terms of this basis, as shown in Table \ref{tab:shapes}.
Shapes constructed from the first 3 basis functions, $\{\mathcal{K}_0,\mathcal{K}_1,\mathcal{K}_2\}$, can alternatively be written in terms of linear combinations of the local, equilateral, and orthogonal templates, which we will refer to as the LEO basis.
For example, the enfolded template can be rewritten in terms of the LEO basis as $S_{enf} = \frac{1}{2}(S_{orth} - S_{equil})$.

As mentioned in \cite{Byun:2013jba}, one advantage of the ${\cal K}_n$ basis is its ability to recover the correct squeezed limits of shapes.
This is especially pertinent for this work, because the similarity between a shape and its template approximation may be more critical for characterizing the model's predictions for the halo bias and halo mass function, since these probes tend to correspond more to the shape's squeezed limit, compared to the CMB bispectrum.
Motivated by this concern, \cite{Senatore:2009gt} constructed an improved orthogonal template, called $S_{orth(2)}$, which in addition to having the same overall features as the original $S_{orth}$, also has the same behavior in the squeezed limit as the primordial shape on which the original orthogonal template was based.
Similarly, \cite{Creminelli:2010qf} constructed a template peaking on enfolded triangles with a vanishing squeezed limit, which we call $S_{enf(2)}$, but arises from higher-derivative operators of single-field models.
Both $S_{orth(2)}$ and $S_{enf(2)}$ can be written exactly in terms of the ${\cal K}_n$ basis, as we show in Table \ref{tab:shapes}.

Furthermore, combinations of the basis modes can create a set of irreducible shapes with the same divergence properties in the squeezed limit \cite{Byun:2013jba}.
For example, we can consider modes that individually diverge by up to $1/k^2$ in the squeezed limit, $\{{\cal K}_0-{\cal K}_6\}$, and,  construct three irreducible templates, in addition to the equilateral template, $S_{equil}$: $\{{\cal K}_{0}+3{\cal K}_{3}-3{\cal K}_{4},  {\cal K}_{2}+2{\cal K}_{3}-2{\cal K}_{5}, 2{\cal K}_{3}-{\cal K}_{6}\}$ that each vanish in the squeezed limit, but have significant differences from $S_{equil}$ away from the squeezed limit.  

A general template that vanishes in the squeezed limit can be reconstructed out of linear combinations of these four templates.
An example of this is the Self-Ordering Scalar Fields (SOSF) model \cite{Figueroa:2010zx}, which while vanishing in the squeezed limit, like the equilateral shape, has principal power in the `aligned'  configuration where $k_1\sim 2 k_2\sim 2k_3$.
In \cite{Figueroa:2010zx}, a fitting function for $S_{SOSF}$ is provided that fits the numerical results to a few percent.
We reconstruct a template, included in Table \ref{tab:shapes}, that has the maximal cosine ($>0.99$) with the fitting function in \cite{Figueroa:2010zx}, and vanishes in the squeezed limit. 
We use the fitting function for the LSS analysis, but use the separable template in Table \ref{tab:shapes} to allow efficient computation of the CMB constraints.

We also consider non-Gaussian shapes that can arise from models with large-scale magnetic fields (which could be generated by vector perturbations during inflation), vector fields that are coupled to the inflaton, and the solid (or elastic) inflation model \cite{Shiraishi:2013vja}. 
The basis describing these, $\{S_0, S_1, S_2\}$ in \cite{Shiraishi:2013vja}, corresponds exactly to a linear combination of $\mathcal{K}_n$ functions, and represents a theoretically-motivated generalization of the local shape, analogous to studies of scale-dependent $f_{NL}$ \cite{Becker:2012yr}.
In this analysis we include consideration of the $S_1$ template used in describing these models, as an example of a shape with an anisotropic squeezed limit, in which their properties differ depending on from which direction, in $k$-configuration space, the squeezed limit is approached.

\begin{table}[!t]
\begin{tabular}{|l|l|}
\hline
Shape & Template in terms of the ${\mathcal K}_n$ basis
\\ \hline
$ S_{local}$  &$\mathcal{K}_2 $
 \\
$ S_{equil}$ & $-2\mathcal{K}_0+6\mathcal{K}_1-3\mathcal{K}_2$
\\
$S_{orth}$  &$-8\mathcal{K}_0+18\mathcal{K}_1-9\mathcal{K}_2$
\\
$S_{enf}$  &  $-3\mathcal{K}_0+6\mathcal{K}_1-3\mathcal{K}_2$
\\ 
$S_{orth(2)}$ &$(1+p)S_{equil} -p\left( \frac{2}{9} {\cal K}_{0}   + \frac{8}{3}{\cal K}_{1} - 2 {\cal K}_{2} + \frac{20}{9} {\cal K}_{3}  - \frac{10}{3} {\cal K}_{4}   + \frac{4}{3} {\cal K}_{5} -  \frac{1}{9} {\cal K}_{6}\right)$
\\
$S_{enf(2)}$ &$(1+\alpha)S_{equil}-\alpha\left(\frac{ 6}{5}{\cal K}_{0}+\frac{16}{5}{\cal K}_{3}-\frac{18}{5}{\cal K}_{4}+\frac{1}{5}{\cal K}_{6}\right)$
\\
${S_{SOSF}}$ &$22.6 \, S_{equil} - 5.98\left( {\cal K}_{0}+3{\cal K}_{3}-3{\cal K}_{4}\right) - 29.5\left(  {\cal K}_{2}+2{\cal K}_{3}-2{\cal K}_{5}\right) + 13.9\left( 2{\cal K}_{3}-{\cal K}_{6}\right)$
\\
$S_{1}$ & $2\mathcal{K}_4-\mathcal{K}_6$
\\ \hline
\end{tabular}
\caption{The ${\mathcal K}_n$ expansion for a variety of templates discussed in the literature. The variables $p$ and $\alpha$, used in $S_{orth(2)}$ and $S_{enf(2)}$, are  chosen to maximize the template's fit to the physical shape. The  separable template for $S_{SOSF}$ is used for the CMB analysis, and is constructed from combinations of the basis functions that vanish in the squeezed limits, with coefficients that maximize the cosine with the fitting function given in  \cite{Figueroa:2010zx}, giving a cosine of $>0.99$. 
 \label{tab:shapes}}
\end{table}

\subsection{Non-Gaussian halo statistics}
\label{subsec:ng_halos}

In general, primordial non-Gaussianity changes the power spectrum of dark matter halos by coupling the local power spectrum to the local long-wavelength matter overdensity. 
This effect was first discovered in N-body simulations of local non-Gaussianity \cite{Dalal:2007cu}, and since then there has been both theoretical and numerical work towards understanding the effect more precisely and how it extends to other types of non-Gaussianity (e.g. \cite{Wagner:2010me,Wagner:2011wx,Schmidt:2010gw,Desjacques:2011jb,Desjacques:2011mq,Scoccimarro:2011pz}).
Here we review how general bispectrum shapes can change the statistics of halos through its effect on the halo bias and halo mass function, adopting the framework based on the peak-background split approach \cite{Scoccimarro:2011pz}. 

Given the primordial gravitational potential, $\Phi(\vec{k})$, where $\Phi = 3\zeta/5$, the dark matter density contrast $\delta(\vec{k})$ at redshift $z$ is described by 
\bea
\delta(\vec{k},z) &=& M(k,z) \Phi(\vec{k}), 
\eea
where
\bea
M(k,z) &\equiv& \dfrac{2 D(z) k^2 T(k)}{3 \Omega_{m,0} H_0^2},
\eea
$D(z)$ is the growth factor normalized to 1 at $z=0$, and $T(k)$ is the matter transfer function.
$M(k,z)$ is straightforward to compute using publicly available codes, such as CAMB\footnote{http://camb.info}.
For simpler notation, from now on we will drop the explicit $z$-dependence in our equations.

In the presence of non-Gaussianity, the matter fluctuations acquire a non-zero skewness, $S_3$, which is an integrated measure of the smoothed matter bispectrum,
\be
S_3(m) \equiv \frac{\la \delta_R^3 \ra}{\sigma_R^4},
\ee
where $\sigma_R^2$ is the variance of the smoothed density contrast,
\be
\sigma_R^2 \equiv \int \dfrac{d^3k}{(2\pi)^3} M(k)^2 W_R(k)^2 P_\Phi(k),
\ee
and the third moment is
\begin{eqnarray}
\la \delta_R^3 \ra &\equiv& \int d^3k \, d^3q \, M(k) M(q) M(|\vec{k}-\vec{q}|) W_R(k) W_R(q) W_R(|\vec{k}-\vec{q}|) B_\Phi(k,q,|\vec{k}-\vec{q}|) \nonumber \\
&=& \frac{2}{(2\pi)^4} \int d{\ln}k \, k^3 M(k) W_R(k) \int d{\ln}q \, q^3 M(q) W_R(q) \int d\mu M(k_\mu) W_R(k_\mu) B_\Phi(k,q,k_\mu). \hspace{1cm}
\end{eqnarray}
$W_R(k)$ is the Fourier transform of a spherical top-hat function of radius $R$, normalized such that $\int dx W_R(x) = 1$, where $R$ corresponds to an effective radius for the halo mass under consideration, with  $R = (3M_{halo}/4\pi \rho_m)^{1/3}$, and $\rho_m$ is the matter density today.
In what follows, we choose to work with the redshift-independent reduced skewness, $\sigma_R S_3(m)$.

The skewness induces changes in the halo mass function, $dn/dm$, that gives the number density, $n$, of halos of mass $m$, and we parametrize the deviation from what the halo mass function would be in the Gaussian case through the factor $R_{NG}$ \cite{LoVerde:2007ri}, defined as
\be \label{eq:Rng}
R_{NG} \equiv \frac{dn_{NG}/dm}{dn_G/dm} =1 + \frac{1}{6} \sigma_R S_3 \left( \nu^3 - 3\nu \right) - \frac{1}{6}\frac{d\sigma_R S_3}{d{\ln}\nu} \left( \nu - \frac{1}{\nu} \right),
\ee
where $\nu \equiv \delta_c / \sigma_R$, and we assume $\delta_c = 1.686$ is the linear threshold for spherical collapse.

In addition to changes in the halo mass function, non-Gaussianity also alters how the halos are distributed with respect to the underlying dark matter fluctuations.
At the linear level, the halo density $\delta_h$ traces the matter fluctuations with a bias factor that depends on the halo mass $m$ and wavenumber $k$, 
\be
\delta_h(m,\vec{k}) \simeq \bh(m,k) \delta(\vec{k}).
\ee
The linear halo bias can then be split up into three contributions: a Gaussian scale-independent term, a non-Gaussian scale-independent term, and a non-Gaussian scale-dependent term,
\be \label{bhalo}
\bh(m,k) = b_G(m) + \Delta b_{si}(m,f_{NL}) + \Delta b_{sd}(m,k,f_{NL}).
\ee
The scale-independent bias correction is derived from $R_{NG}$ as \cite{Desjacques:2008vf}
\begin{eqnarray} \label{eq:bsi}
\Delta b_{si}(m,f_{NL}) &\equiv& -\frac{1}{\sigma_R} \frac{\partial {\ln}R_{NG} }{\partial\nu} \nonumber \\
&=&
 -\frac{1}{6\sigma_R} \frac{1}{R_{NG}} 
\left[ 3\sigma_R S_3\left(\nu^2-1\right) 
+ \frac{d\sigma_R S_3}{d{\ln}\nu} \left(\nu^2 - 4 - \frac{1}{\nu^2}\right)  
- \frac{d^2\sigma_R S_3}{d{\ln}\nu^2} \left( 1-\frac{1}{\nu^2} \right) \right]. \hspace{1cm}
\end{eqnarray}

The scale-dependent term $\Delta b_{sd}(m,k)$ is given by 
\be \label{eq:bsd}
\Delta b_{sd}(m,k,f_{NL}) = \dfrac{\delta_c[b_G(m) + \Delta b_{si} -1]}{2M(k)} \dfrac{I_{21}(m,k)}{\sigma_R^2} + \dfrac{1}{M(k)} \partial_{\ln \sigma_R^2} \left[\dfrac{I_{21}(m,k)}{\sigma_R^2}\right].
\ee
$I_{21}$ is defined and calculated as
\begin{eqnarray}
I_{21}(m,k) &\equiv \frac{1}{P_\Phi(k)}
	\int \frac{d^3q}{(2\pi)^3} \; M(q) W_R(q) M(|\vec{k}-\vec{q}|) W_R(|\vec{k}-\vec{q}|) B_\Phi(k,q,|\vec{k}-\vec{q}|) \nonumber \\
&= \frac{1}{(2\pi)^2 P_\Phi(k)}
	\int d{\ln}q \; q^3 M(q) W_R(q) \int_{-1}^{1} d\mu \; M(k_\mu) W_R(k_\mu) B_\Phi(k,q,k_\mu), \label{eq:I21}
\end{eqnarray}
where by construction, $\vec{k_\mu} = \vec{k} - \vec{q}$ such that $k_\mu^2 = k^2 + q^2 - 2kq\mu$, and $\mu$ is the cosine of the angle between $\vec{q}$ and $\vec{k}$.  This multidimensional integral and others in this work are computed using Cuba, a publicly available code for numerical integration of multidimensional integrals\footnote{http://www.feynarts.de/cuba/}.

In the local case, $I_{21}/\sigma_R^2$, the first term in the scale-dependent bias, tends towards a constant value of $4 f_{NL}$ on large scales, such that the contribution from the second, derivative term, $\partial_{\ln \sigma_R^2} [ I_{21}(m,k)/ \sigma_R^2]$ is negligible.
However, in general the inclusion of both terms is important: the first and second terms typically have opposing signs and partially cancel each other on scales above a few Mpc. 
For the templates where the two terms are comparable, such as in the equilateral and orthogonal cases, the large scale, $k\rightarrow 0$ scaling behavior and small scale properties are determined by the cancellation.
The relative weight of each term changes with the halo mass, and gives rise to mass-dependent biases. 

When computing $\Delta b_{sd}$ on large scales, the integral over $\mu$ in \eqref{eq:I21} averages over squeezed configurations. 
For isotropic shapes, the absence of  $\mu$ dependence in the leading term in the squeezed limit expansion of the shape in powers of $k/q$ gives a trivial mapping of the large scale scale-dependence of the large-scale halo bias. 
For shapes for which the squeezed limit is anisotropic in $\mu$, the $\mu$ integral can alter the sign and amplitude of the leading $k$ scaling in the bias, or make it vanish. 
Thus interpreting the halo bias data using just large scale data, and presuming that the primordial shape is isotropic, could potentially lead to incorrect or incomplete conclusions about the shape determination, and miss an important sector of the theory space.

\subsection{The halo model}
\label{subsec:halomodel}

In this section, we first briefly summarize our implementation of the halo model for computing non-linear matter and galaxy power spectra (see \cite{Cooray:2002dia} for a detailed review of the halo model), and then describe our resulting power spectra from different non-Gaussian shapes.

In the halo model, the halo matter power spectrum $P_h$ is the sum of correlations between separated halos, $P^{2h}$, and within the same halo, $P^{1h}$, $P_h(k) = P^{1h}(k) + P^{2h}(k)$, with
\begin{eqnarray}
P^{1h}(k) &=& \int dm \frac{dn}{dm} \left( \frac{m}{\bar{\rho}} \right)^2 |u(k|m)|^2, \\
P^{2h}(k) &=& \left[ \int dm \frac{dn}{dm} \left( \frac{m}{\bar{\rho}} \right) u(k|m) b_h(m,k) \right]^2 P_{linear}(k).
\end{eqnarray}
Here $P_{linear}$ is the linear dark matter spectrum, $dn/dm$ is the halo mass function, and $b_h$ is the halo bias.
In the fiducial Gaussian case, we compute $dn^G/dm$ and $b_G$ using the numerical fits in \cite{Pillepich:2008ka}.
We include non-Gaussianity in the halo mass function using the $R_{NG}$ parameter, and  in the halo bias, parametrized by $\Delta b_{si}$ and $\Delta b_{sd}$, as described in section \ref{subsec:ng_halos}.
$u(k|m)$ is the Fourier transform of the halo density profile that describes the distribution of matter within each halo.
We assume the Navarro-Frenk-White (NFW) profile \cite{Navarro:1995iw,Navarro:1996gj}, which has two free parameters that can be chosen as a scale radius, $r_s$, and a corresponding matter density, $\rho_s$.
However, it is also common to choose the two free parameters to be how the virial radius, $R_{vir}$, of the halo is defined, and how the concentration of each halo, $c$, depends on the halo mass and redshift.
Here we define $R_{vir}$ such that it contains an average density that is $\Delta_{vir}=200$ times the average matter density of the universe, and use the concentration from \cite{Bullock:1999he}.
Our choices of these inputs to the halo model calculation were made to facilitate comparisons between our results and those of previous forecasts.
In particular, we have used the same Gaussian halo mass function and halo density profile that were noted by \cite{Giannantonio:2011ya} to provide good agreement with simulations.

Implementing the halo model requires two additional constraints to be imposed by hand: (1) that all matter is contained in halos, and (2) that the calculated Gaussian matter power spectrum on large scales matches the linear expectation, $P_{linear}(k)$.  The numerical details of enforcing these constraints are described in \cite{Fedeli:2009mt}.

\subsection{The galaxy bias model}
\label{subsec:galmodel}
One can compute the galaxy power spectrum, $P_{gal}(k)$, by using theoretical and empirical prescriptions for how galaxies are arranged within halos of a given mass. With assumptions on the distribution of galaxies within the halo, $u_{gal}$, and the moments of the statistical distribution of the number of galaxies with a halo of a given mass, the Halo Occupation Distribution (HOD), $\left<N_{gal}|m\right>$, the galaxy power spectrum can be written:
\begin{eqnarray}
P^{1h}_{gal}(k,z) &=& \int dm \frac{dn}{dm}\frac{ \la N_{gal}(N_{gal}-1)|m \ra}{\bar{n}_{gal}^2} |u_{gal}(k|m)|^p, \\
P^{2h}_{gal}(k,z) &=& \left[ \int dm \frac{dn}{dm} \frac{ \la N_{gal}|m \ra}{\bar{n}_{gal}} u_{gal}(k|m) b_h(m,k) \right]^2 P_{linear}(k,z),
\end{eqnarray}
where
\be
\bar{n}_{gal}(z) = \int dm \frac{dn}{dm} \la N_{gal}|m \ra.
\ee
We make the reasonable assumption that galaxies are distributed within halos in a similar way that dark matter particles are, i.e. $u_{gal}(k|m) = u(k|m)$, and take the, redshift independent, moments $\la N_{gal}|m \ra$ and $\la N_{gal}(N_{gal}-1)|m \ra$ of the halo occupation distribution used in \cite{Fedeli:2010ud}.
$p$ in the $P^{1h}$ term is either 1 or 2, depending on whether $\la N_{gal}(N_{gal}-1)|m \ra < 1$ or not.

As the mapping between galaxies and the underlying dark matter, the galaxy bias, $b_{gal}(k,z)=\sqrt{P_{gal}/P_{h}}$, can be probed in different ways -- observationally, theoretically, and with simulations. In the large scale limit, $u_{gal}\approx 1$ and the 2-halo term dominates so that
\be
b_{gal} (z)\approx \int dm \frac{dn}{dm}
 \frac{\la N_{gal}|m \ra}{\bar{n}_{gal}}b_h(m,k\ll1).
\ee

\begin{figure}[t]
\centering
\includegraphics[width=0.5\textwidth]{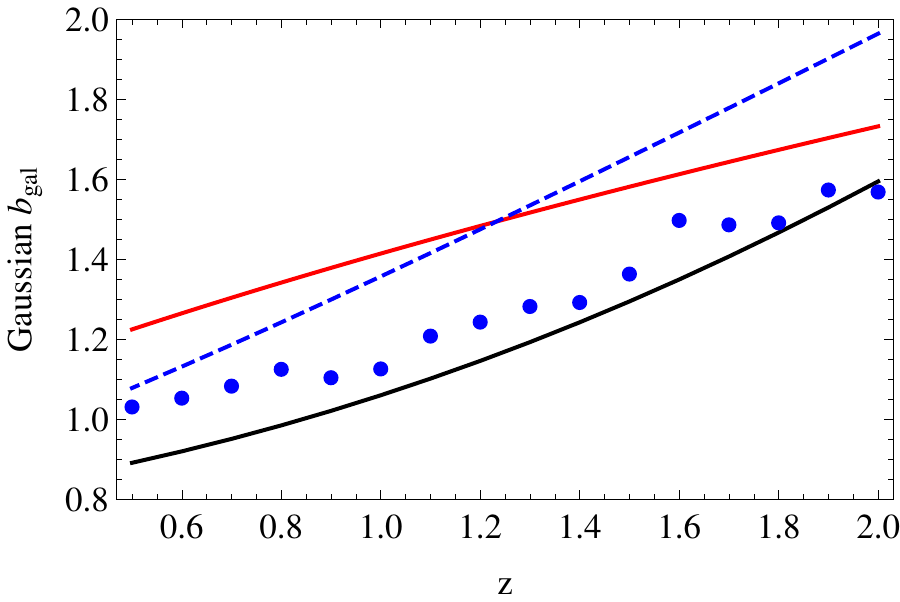}
\caption{Fiducial Gaussian galaxy biases in this work from \cite{Orsi:2009mj} [blue circles] and computed from the halo model we consider for ELGs [black].  
Other bias parametrizations in the literature are shown for comparison:  
$b_{gal}^G(z) = \sqrt{1+z}$ [red] \cite{Giannantonio:2011ya}, and ELG bias $b_{gal}^G(z) = 0.84/D(z)$ [blue dashed] \cite{Levi:2013gra}.}
\label{fig:gaussian_bgal_vs_z}
\end{figure}

As our fiducial, Gaussian, galaxy bias, we use the values from semi-analytic models of galaxy formation from Orsi et al. \cite{Orsi:2009mj}, which are in broad agreement with the range of fiducial Gaussian bias values that are taken in the literature for different comparable forecasts. 
Figure \ref{fig:gaussian_bgal_vs_z} gives a comparison of the halo model Gaussian galaxy bias evaluated for halos consistent with emission line galaxies (ELGs) and the Orsi et al. paper, along with others assumed in the literature.

The galaxy bias can be measured most directly if the galaxies and matter are observed in overlapping fields, for instance, with galaxy-lensing cross correlations.
Such constraints have been analyzed while assuming the Gaussian case \cite{Hoekstra:2001pi,Hoekstra:2002gs,Seljak:2004sj,Cacciato:2012gv,Jullo:2012ty,More:2014uva}, as well as for simultaneously constraining templates of non-Gaussianity \cite{Jeong:2009wi,Takeuchi:2010bc,Giannantonio:2011ya,Giannantonio:2013kqa}.
In observations without a lensing component, the matter field is less directly inferred from the data, due to the complex processes of galaxy formation.
However, we know from theoretical and observational considerations that galaxy formation is strongly correlated with their host halos, which allows for full encapsulation of complex formation processes into a HOD with a few free parameters describing the number and spatial distribution of galaxies inhabiting a halo of a given mass \cite{Jing:1997nb,ShethSaslaw94}.
The HOD allows galaxy survey measurements to be more easily related to the underlying halos, the statistics of which are more directly predicted by theoretical cosmology-dependent parameters.
With the HOD as a tool to bridge the gap between galaxies and halos, other probes beyond  galaxy-lensing correlations have been employed to infer the galaxy bias.
One such technique is to employ dark matter simulations and a halo finder algorithm, which can be done for both Gaussian and non-Gaussian initial conditions, to see how halo power spectrum and bispectrum statistics are altered by primordial shapes \cite{Wagner:2010me,Wagner:2011wx,Sefusatti:2011gt}.
Given the output of such simulations of halos, the HOD can further these predictions to the level of galaxies, allowing for easier comparison of theories with data.
For example, higher-order galaxy correlations, such as the bispectrum of galaxies, can also constraint galaxy bias parameters in the Gaussian case \cite{Verde:2001sf,Marin:2013bbb,Gil-Marin:2014sta}, and the bias parameters simultaneously with the $f_{NL}$ parameters in non-Gaussian cosmologies \cite{Sefusatti:2007ih}.
While a detailed description and implementation of such techniques is outside the scope of this paper, the combination of these studies paints a picture of steady improvements in the theoretical understanding and observational measurements of the galaxy bias, and understanding how to extract non-Gaussian constraints simultaneously.   

\subsection{Fisher matrix approach for LSS and CMB}
\label{subsec:fisher_approach}

We conduct a Fisher matrix analysis as a preliminary tool to estimate potential constraints on bispectrum shapes with upcoming surveys. 
This approach has been used in a number of previous analyses that also aim to extract constraints on primordial NG from the halo bias \cite{Giannantonio:2011ya,Sefusatti:2012ye,Becker:2012yr,Agarwal:2013qta,Shandera:2010ei}.
The analysis we present here assumes a spectroscopic galaxy survey from Euclid-like and DESI-like experiments, where the observed galaxy power spectrum can be used to infer the underlying dark matter distribution, which in turn contains traces of primordial non-Gaussianity.
In doing so, however, we must account for the various astrophysical and observational effects that can change how the observed galaxies are related to the halo model's theoretically predicted galaxy power spectrum.

We model the relation between the theoretical galaxy (position) power spectrum we compute from the halo model, $P_{gal}$, and the one that is observed, $P_{gal}^{obs}$, as
\be \label{eq:obs P_g}
P_{gal}^{obs}(k,\tilde{\mu},z) = 
		\left[1+ \frac{f(z) \tilde{\mu}^2}{b_{gal}(k,z) }\right]^2 P_{gal}(k,z)
		e^{- \mathcal{D}(k,\tilde{\mu},z)}
\ee
where $\tilde{\mu}$ is the cosine of the angle between the wavenumber $k$ and the line of sight, and should not be confused with $\mu$ in the definition of the non-Gaussian shape.  
The bracketed factor in \eqref{eq:obs P_g} uses the Kaiser formula \cite{Kaiser:1987qv} to model the effect of redshift-space distortions (RSD), and we take $f \equiv \Omega_m(z)^{0.55}$ as the linear growth rate.
The exponential damping, ${\mathcal D}$, contains contributions from the redshift uncertainty of the survey ($\sigma_z/z = 0.001$), peculiar velocities of galaxies, and the nonlinear growth of structure.
We model this component as in \cite{Seo:2007ns}, and choose to set a maximum wavenumber at $k_{max} = 0.25 \, h/\mathrm{Mpc}$ and a non-linear damping factor of $p_{NL} = 0.5$, corresponding to the assumption that our knowledge of non-linear scales is sufficient to remove half of the contamination from non-linear growth of structure through reconstruction \cite{Padmanabhan:2012hf}.

The set of parameters to be constrained, $\vec{\theta}$,  includes the non-Gaussianity parameters we wish to constrain, as well as nuisance parameters that we wish to marginalize over.
We allow for the potential for additional, external information about the galaxy bias in the analysis by allowing the Gaussian galaxy bias in each redshift to act as a nuisance parameter, that can be marginalized over with varying priors.
For this analysis, our parameter set is $\vec{\theta} = \{ f_{NL}; b_{gal}^G(z_1), ..., b_{gal}^G(z_N) \}$,
where the Gaussian galaxy bias parameters, $b_{gal}^G$, for each of the $N$ $z$-bins are included. We consider scenarios in which they are fully marginalized over, representing no additional information about the bias from external analyses, to perfect knowledge, with no marginalization, and gradations in between through the imposition of independent, Gaussian priors on each.

The Fisher matrix contribution in a given redshift slice, now including the sky fraction, is given by
\be \label{eq:Fz}
	F_{\alpha\beta}(z) = \frac{f_{sky}}{8\pi^2} 
		\int_{\ln k_{min}}^{\ln k_{max}} k^3 \; dlnk \int_{-1}^{1} d\tilde{\mu}
		\frac{\partial \ln P_{gal}^{obs}(k,\tilde{\mu},z)}{\partial \theta_\alpha}
		\frac{\partial \ln P_{gal}^{obs}(k,\tilde{\mu},z)}{\partial \theta_\beta} 
		V_{eff}(k,\tilde{\mu},z),
\ee
and evaluated at the fiducial cosmology, for which we use the best-fit cosmological parameters from the WMAP9 results \cite{Bennett:2012fp}, assuming no non-Gaussianity: 
$\Omega_bh^2 = 0.02264$, $\Omega_ch^2 = 0.1138$, $\Omega_\Lambda = 0.721$, $\Delta_\mathcal{R}^2 = 2.41\times 10^9$, $n_s = 0.972$, $\tau = 0.089$.
The effective inverse data covariance is given by 
\begin{eqnarray}
	V_{eff}(k,\tilde{\mu},z) &=& V_{survey}(z) \left[ \frac{\bar{n}_{gal}^{obs}(z) P_{gal}^{obs}(k,\tilde{\mu},z)}{1+\bar{n}_{gal}^{obs}(z) P_{gal}^{obs}(k,\tilde{\mu},z)} \right]^2 \\
	V_{survey}(z) &=& \frac{4 \pi}{3} \left[ \chi(z_{max})^3 - \chi(z_{min})^3 \right],
\end{eqnarray}
where $\chi$ is the comoving distance, $\bar{n}_{gal}^{obs}$ is the predicted number density of galaxies observed for the survey in the redshift bin, and $z_{max}$ and $z_{min}$ are the boundary redshifts for a given $z$-bin.
To combine the constraints from multiple redshift slices and get the total Fisher error, we simply add the Fisher matrices from each redshift bin.

For the Euclid-like forecast, we consider 27 redshift bins from $0.7 < z < 2$, with bin width $\Delta z = 0.05$, where the number densities of galaxies within each bin are taken from \cite{Wang:2012bx} and $f_{sky} = 20,000 \; \mathrm{deg}^2$.
For the DESI-like forecast, the constraints are calculated separately for the ELG and LRG galaxy populations, before being combined.
The DESI ELG constraint uses 11 redshift bins from $0.6 < z < 1.7$ with $\Delta z = 0.1$, while the DESI LRG constraint uses 6 redshift bins from $0.6 < z < 1.2$ with $\Delta z = 0.1$.
For DESI, the estimated number densities of LRG and ELG galaxies is as per Table 2.3 of the DESI Conceptual Design Report \cite{DESI:Sep14}, and assumes a $14,000 \; \mathrm{deg}^2$ survey.

The $k$-dependence of the inverse covariance matrix of $\ln P_{gal}^{obs}$ is proportional to $k^3 V_{eff}$ which scales roughly as $\propto k^3$. 
The mildly non-linear scales, where both the halo mass function and scale-dependent bias are important and sensitive to a broad range of primordial shape configurations, are potentially better measured than larger scales.
The additional effect of the bias marginalization has to be taken into account, however, in order to assess the projected constraints.

Where the large scale structure constraints are combined with CMB constraints, we have followed the CMB Fisher calculation described in \cite{Byun:2013jba}, except using the fiducial background cosmology consistent with this work.

%% file: sept18_halobiaspaper_results.tex
\section{Findings}
\label{sec:results}

In this section we present the major findings. Sections \ref{subsec:squeezed} and \ref{subsec:config} describe, analytically and numerically, how the halo bias and halo mass function probe different regions of the primordial shape.
Section \ref{subsec:fisher} presents the forecasted constraints on fixed shapes, as well as constraints for the distinguishability between shapes and general $k$-configuration dependent constraints.

\subsection{Squeezed limit behavior}
\label{subsec:squeezed}

\begin{table}[!t]
\centering
\begin{tabular}{|l|l|}
\hline
Shape & \multicolumn{1}{|l|}{Squeezed limit form to $\mathcal{O}(k/q)$}
\\ \hline
$S_{local}$  & $\dfrac{2}{3}\dfrac{q}{k} \hspace{0.2cm} - \dfrac{\mu}{3} \hspace{1.35cm} + \dfrac{1}{6}(1+3\mu^2)\dfrac{k}{q}  $
\\
$S_{orth}$   & \hspace{0.75cm} $- 2 \hspace{1.55cm} + 6(1-\mu^2)\dfrac{k}{q}$
\\
$S_{enf}$ &\hspace{0.9cm}$ -1   \hspace{1.6cm} +2(1-\mu^2)\dfrac{k}{q}$
\\
$S_{equil}$ & \hspace{3.3cm} $2(1-\mu^2)\dfrac{k}{q} $
\\
$S_{ortho(2)}$  & \hspace{3.3cm} $2\left(1-\mu^2 - \alpha\left(\dfrac{5}{27}+\mu^2\right)\right)\dfrac{k}{q}$
\\
$S_{enf(2)}$ &  \hspace{3.0cm} $- 2\left(1-\mu^2-\dfrac{16}{9}\alpha\right)\dfrac{k}{q}$
\\
$S_{1}$ &   \hspace{1.05cm}$\dfrac{2}{3}(1 - 4\mu^2)$ + $\dfrac{8}{3}\mu(1-\mu^2)\dfrac{k}{q}$
\\ 
${S_{SOSF}}$ & \hspace{3.20cm}$ 13.3 \dfrac{k}{q}$
\\ \hline
\end{tabular}
\caption{The squeezed limit expansion, in powers of $k/q$  as $k/q\rightarrow0$, for a variety of templates described in section \ref{subsec:templates}. 
\label{tab:sq}}
\end{table}

On large scales where $k \rightarrow 0$, the halo bias in \eqref{eq:bsd} is a weighted average over squeezed limit configurations of the primordial shape, i.e. configurations with $k/q \ll 1$ with $\mu$ a free parameter. Table \ref{tab:sq} shows the squeezed limit forms for the templates discussed in section \ref{subsec:templates}. 

Many shapes discussed in the literature,  such as the local, orthogonal and enfolded shapes, are isotropic, meaning the leading order term has no $\mu$ dependence as $k/q \rightarrow 0$.  For an isotropic shape, the shape's squeezed limit is $\mu$-independent, so any value of $\mu$ can be selected to simplify the calculation of the shape's corresponding scale-dependent halo bias on large scales. Typically the squeezed isosceles configuration is chosen, equivalent to taking $q = k_\mu$, such that $\mu=k/2q\rightarrow 0$, to show that for a primordial shape which scales like $\propto 1/k^n$ in the squeezed limit, the scale-dependent halo bias will be $\propto 1/k^{n+1}$.
For example, in the local case which diverges as $S_{local}\propto 1/k$ in the squeezed limit, it is straightforward to show that $I_{21}(m)/\sigma_R^2 \rightarrow 4 f_{NL}$ as $k \rightarrow 0$, such that for a fixed halo mass, the halo bias on large-scales is $\Delta b_{sd} \propto 1/M(k)$, resulting in a strongly scale-dependent bias, $\Delta b_{sd} \propto 1/k^2$. 
Even though for isotropic shapes the large-scale behavior of $\Delta b_{sd}$ has a clear interpretation of the shape's squeezed limit, the same is not true of anisotropic shapes, which may have an unusual variety of $\mu$-dependent limits while still appearing degenerate with other isotropic or anisotropic shapes.
Considering the information from the halo bias on small scales may help to disentangle the two.

The $\mathcal{K}_n$ basis also allows for a general $\mu$-dependence, beyond that of the local-like cases considered in \cite{Shiraishi:2013vja}.
Indeed, one can even employ the basis to purposely construct a general shape which is such that the averaging over squeezed configurations produces an unexpected halo bias which breaks the rule of thumb.
For example, it is possible to construct an anisotropic shape which goes to a constant in the primordial squeezed limit, similarly to $S_1$, while having a $\mu$ dependence such that the large-scale halo bias is scale-independent.
This may be a point of concern if a scale-independent bias is confirmed through future surveys:
The single-field consistency relation states that single-field models of inflation should vanish in the primordial squeezed limit.  
Then using the rule of thumb, one would expect a lack of a scale-dependent bias detection to confirm inflationary models of single-field origin.
However, such data may still be consistent with general anisotropic models which do not satisfy the consistency relation, but still yield scale-independent halo biases.

\begin{figure}[!t]
\centering
\includegraphics[width=\textwidth]{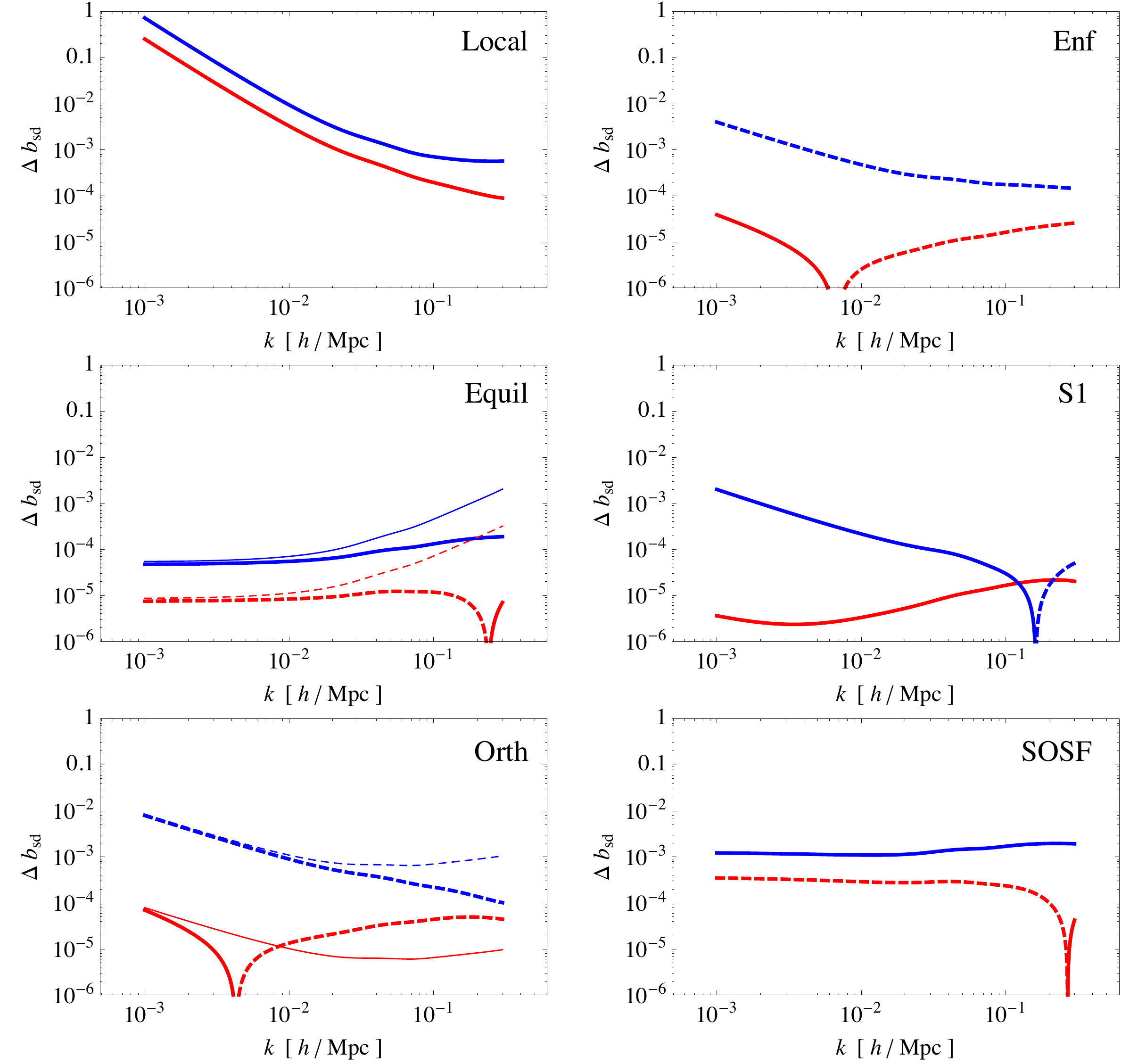}
\caption{
The left panels, from top to bottom, shows the scale-dependent halo bias using the analytic derivation  [thick], as in (\ref{eq:bsd}), and the asymptotic form \cite{Schmidt:2010gw} if extrapolated to small scales [thin],  for two halo masses, $10^{12.5}M_\odot/h$ [red] and $10^{13.5}$ $M_\odot/h$ [blue], for the local, orthogonal, and equilateral shapes at $z=1$ for $f_{NL}=1$. The right panels, from top to bottom, show the scale-dependent bias, for the same halo masses, for the enfolded, anisotropic, and SOSF shapes, respectively. Dashed curves denote negative values.}
\label{fig:bsd}
\end{figure}

Taking the $k \rightarrow 0$ limit of \eqref{eq:bsd} yields a good power-law approximation to a shape's bias on large scales, however, in general as $k$ increases towards smaller scales the bias takes on a more complex scale-dependence.
In this work, since we include both large and small scale power spectra in our Fisher forecasts, 
we use the full $k$-dependence of the bias from a numerical calculation of \eqref{eq:bsd} and \eqref{eq:I21} for each $k$.
In other analyses, small scale behavior has been approximated by extending the large scale-derived, asymptotic, expressions for the halo bias, using an analytically derived $k\rightarrow0$ expression for $I_{21}(m,k)/\sigma_R^2$ and its derivative in evaluating \eqref{eq:bsd}.

Figure \ref{fig:bsd} compares the fully $k$-dependent and asymptotic forms of the bias for the local, equilateral, and orthogonal shapes and shows the differences between the two results on small scales for the orthogonal and equilateral templates.
The right hand panels shows the differences in large scale and small scale properties, and halo mass dependence, for the isotropic shapes, $S_{enf}$ and $S_{SOSF}$, and anisotropic shape, $S_{1}$.
The figure also shows that including information on mildly non-linear scales, $\sim 10$ Mpc, can reveal richer information about the shape than purely considering the large-scale bias.
While two shapes can have the same scaling (up to a constant) in the squeezed limit, and thus they both have a scale-dependent bias $\propto 1/k^n$ on large scales, on smaller scales, and at different halo masses, their bias functions can differ.

We note that particularly in the equilateral template constraints, the use of an asymptotic approximation for the scale dependent bias, $\Delta b_{sd}$, that extrapolates the large scale behavior to small scales, leads to tighter constraints on $f_{NL}$ \cite{Giannantonio:2011ya}. This is due to an artificial increase in the amplitude of the bias on small scales relative to when the full analytic expression for the halo bias is used.

\subsection{Configurations probed by the halo bias and mass function}
\label{subsec:config}

To explore which configurations of the shape the scale-dependent bias probes best, and allow for the consideration of shapes that are anisotropic  in $\mu$ as one approaches the squeezed limit, we define a weight function, $w_{sd}$, that encodes the relative importance to the bias per unit shape amplitude of each ${\ln} q$ and $\mu$, for a tracer of halo mass $m$ and wavenumber $k$: 
\be
\Delta b_{sd}(m,k) = \int d{\ln}q \int d\mu \; w_{sd}(k,{\ln}q,\mu,m) S(k,q,k_\mu). 
\label{eq:bsdweight}
\ee

\begin{figure}[!t]
    \centering
    \begin{subfigure}{0.45\textwidth}
    \centering
        \includegraphics[width=\textwidth]{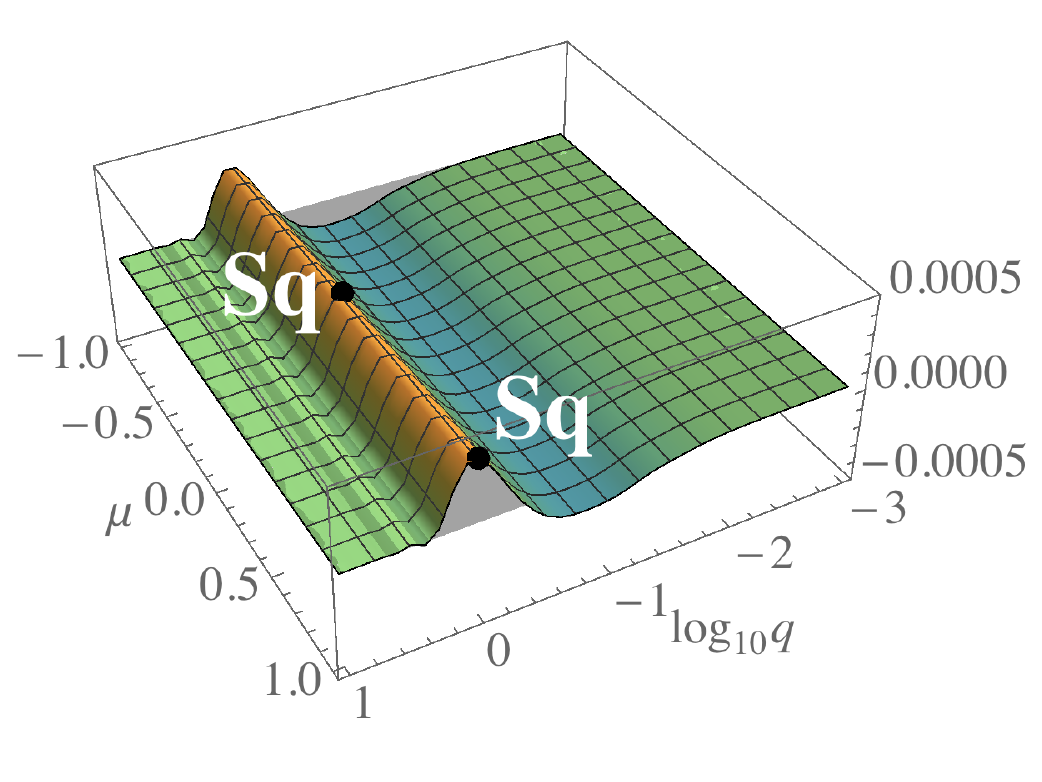}
        \label{fig:bsd_weight_m_12pt5_k_0pt001}
        \caption{$m = 10^{12.5} \, M_\odot/h$, $k = 10^{-3} \, h/\mathrm{Mpc}$}
    \end{subfigure}%
    \begin{subfigure}{0.45\textwidth}
    \centering
        \includegraphics[width=\textwidth]{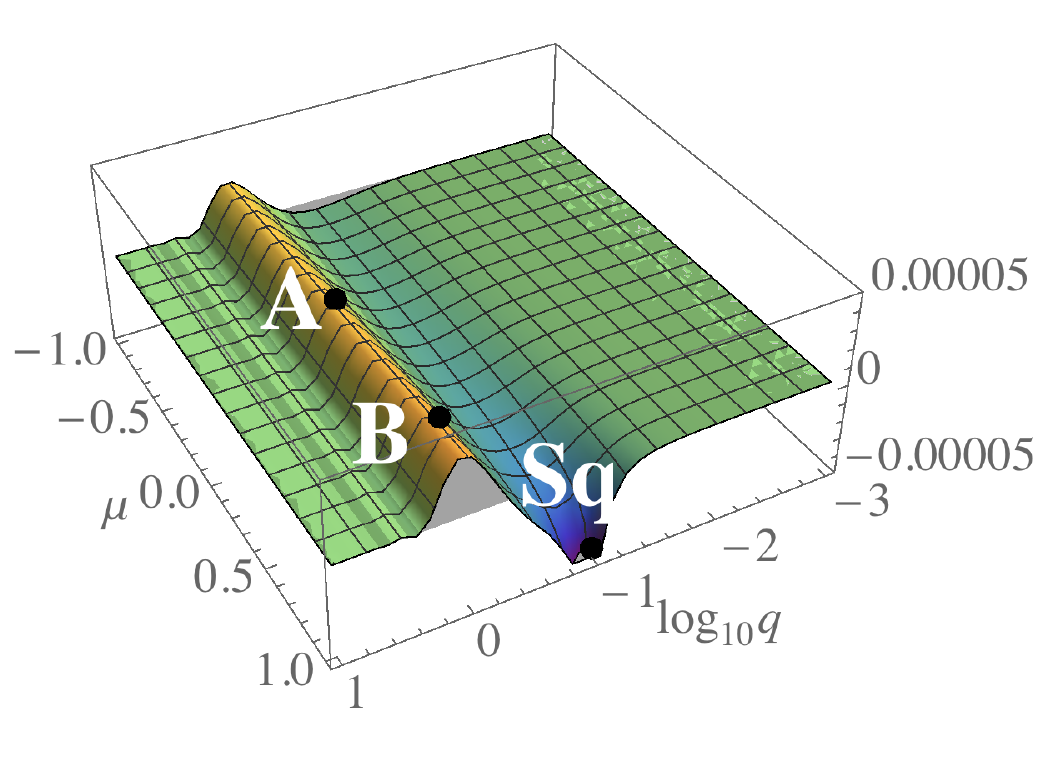}
        \label{fig:bsd_weight_m_12pt5_k_0pt1}
        \caption{$m = 10^{12.5} \, M_\odot/h$, $k = 0.1 \, h/\mathrm{Mpc}$}
    \end{subfigure}
\\
    \begin{subfigure}{0.45\textwidth}
    \centering
        \includegraphics[width=\textwidth]{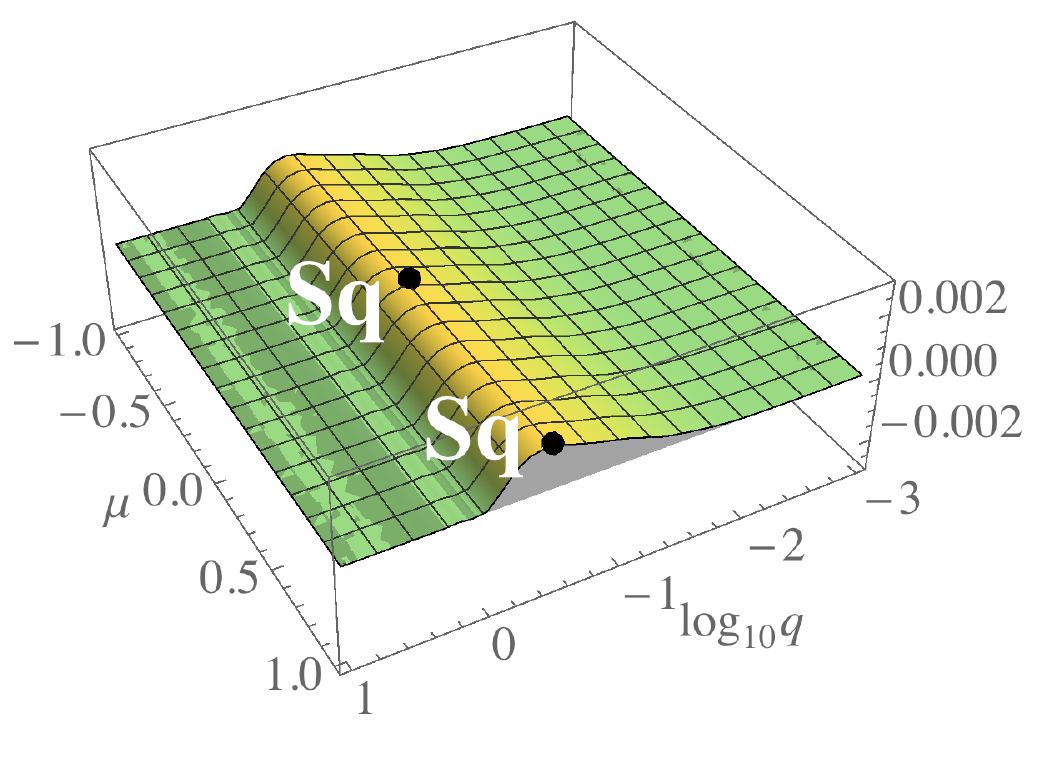}
        \label{fig:bsd_weight_m_13pt5_k_0pt001}
         \caption{$m = 10^{13.5} \, M_\odot/h$, $k = 10^{-3} \, h/\mathrm{Mpc}$}
    \end{subfigure}%
    \begin{subfigure}{0.45\textwidth}
    \centering
        \includegraphics[width=\textwidth]{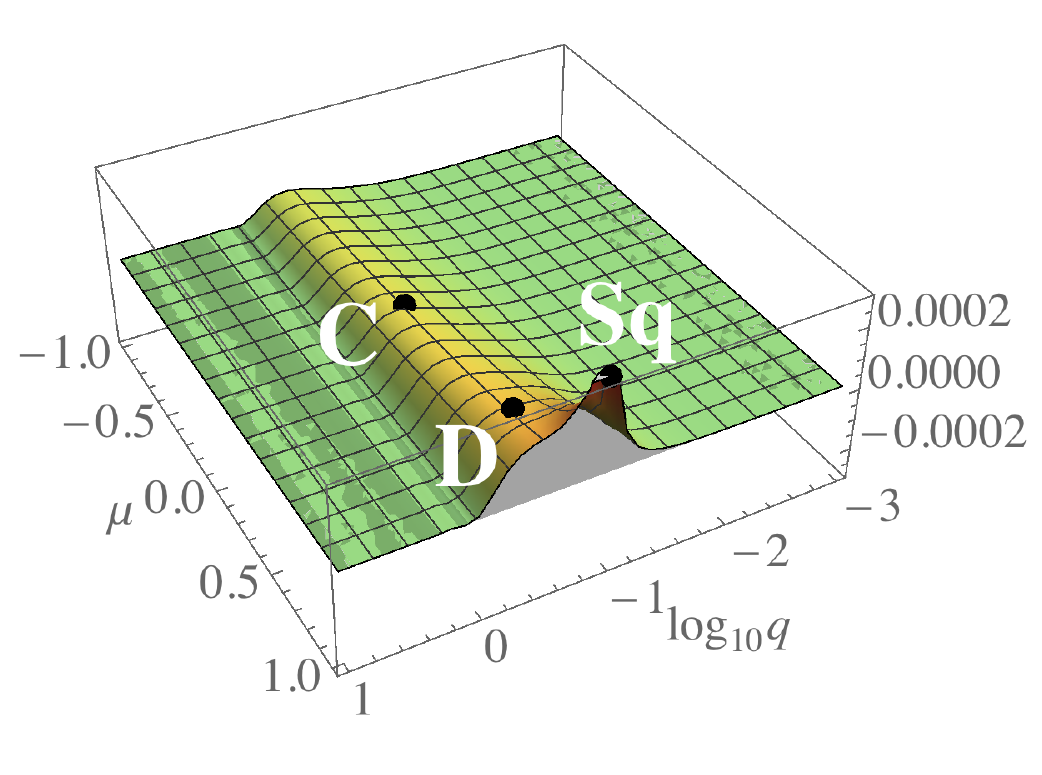}
        \label{fig:bsd_weight_m_13pt5_k_0pt1}
        \caption{$m = 10^{13.5} \, M_\odot/h$, $k = 0.1 \, h/\mathrm{Mpc}$}
    \end{subfigure}
\\
    \begin{subfigure}{0.25\textwidth}
    \centering
        \includegraphics[width=\textwidth]{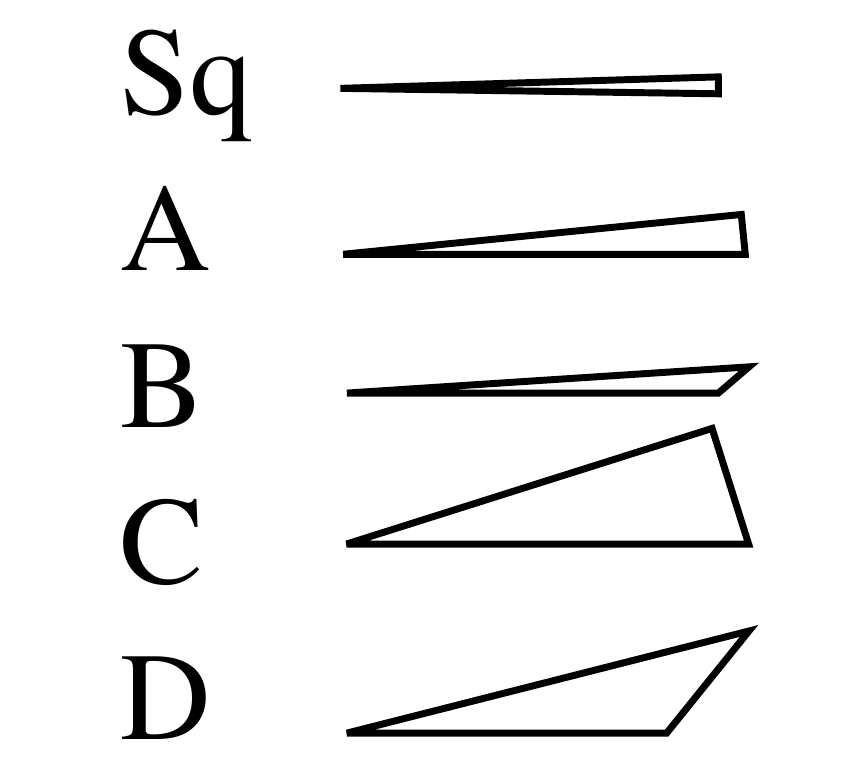}
        \label{fig:bsd_triangles}
        \caption{}
    \end{subfigure}
\caption{The halo bias weight function, $w_{sd}$, for halo masses $m = 10^{12.5} M_\odot/h$ [top] and $m = 10^{13.5} M_\odot/h$ [middle], and scales $k = 10^{-3} \, h/\mathrm{Mpc}$ [left] and $k = 0.1 \, h/\mathrm{Mpc}$ [right] as a function of the wavenumber $q$ and angle $\mu$ used to described the bispectrum shape.
We have labeled specific configurations at which each weight shown peaks.
Some points are simply labeled as `Sq' signifying that the triangular configuration is significantly squeezed and not easily illustrated to scale.
For weights at $k = 0.1 \, h/\mathrm{Mpc}$, we label other configurations and illustrate the corresponding triangles, which are much less squeezed than those labeled `Sq', showing that $\Delta b_{sd}$ on small scales can probe configurations beyond very squeezed triangles.
}
    \label{fig:bsdweight}
\end{figure}

Figure \ref{fig:bsdweight} shows how the precise behavior of the weight function for the scale-dependent bias is dependent upon the halo mass $m$ and the triangular configuration formed by the $(k,q,k_\mu)$ wavenumbers.
On very large scales, $k \sim 0.001 \, h/\mathrm{Mpc}$, the bias primarily probes squeezed configurations for which $k/q \lesssim 0.01$, while on small and mildly non-linear scales of $k \sim 0.1 \, h/\mathrm{Mpc}$, the bias probes a wider range of configurations, including those that are not significantly squeezed.
On intermediate scales (not shown), $k \sim 0.01 \,h/\mathrm{Mpc}$, the weight rises for configurations that are moderately squeezed, $k/q \sim 0.01-0.1$, with an overall amplitude comparable to the weight function of the $k \sim 0.1 \,h/\mathrm{Mpc}$ slice, for the same halo mass.
In all cases, the weights do not pick out a single, preferred value of $\mu$, but are equally sensitive to all values of $\mu$ for a given value of $k$ and $q$.

\begin{figure}[h]
\centering
        \includegraphics[width=0.5\textwidth]{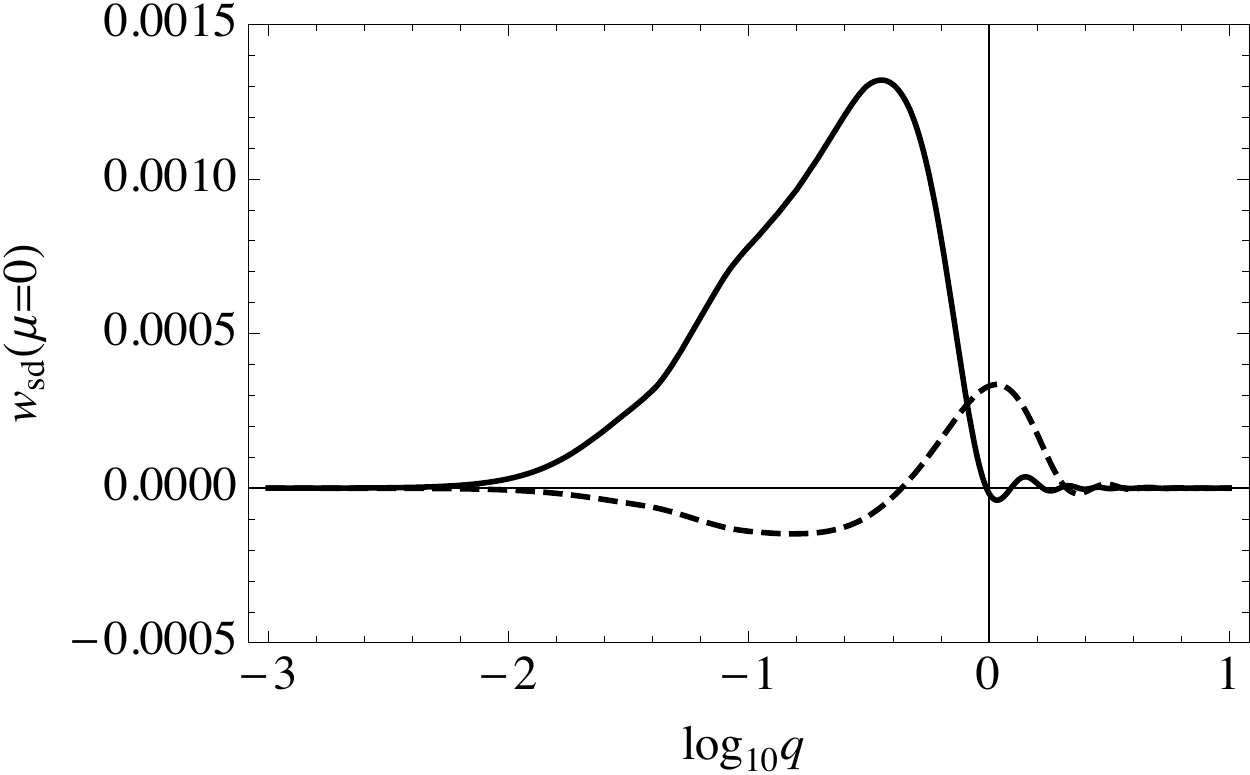}
    \caption{The solid and dashed curves respectively show the $\mu=0$ slice of the weight from the scale-dependent bias for $k = 10^{-3} \, h/\mathrm{Mpc}$ and $M = 10^{13.5} M_\odot/h$ and $M = 10^{12.5} M_\odot/h$ cases.}
    \label{fig:bsd weight mu eq 0 slice}
\end{figure}

To show the mass-dependence in another way,  in Figure \ref{fig:bsd weight mu eq 0 slice} we show just the $\mu = 0$ slice of $w_{sd}$ for $k = 0.001 \, h/\mathrm{Mpc}$ and two halo masses.
The figure illustrates that while larger halo masses have a weight function that is broader and larger in amplitude than for less massive halos suggesting that measurements across a range of halos may yield complementary information by weighting the halo bias to the primordial shape in a different way.

\begin{figure}[!h]
\centering
    \begin{subfigure}{0.5\textwidth}
    \centering
        \includegraphics[width=\textwidth]{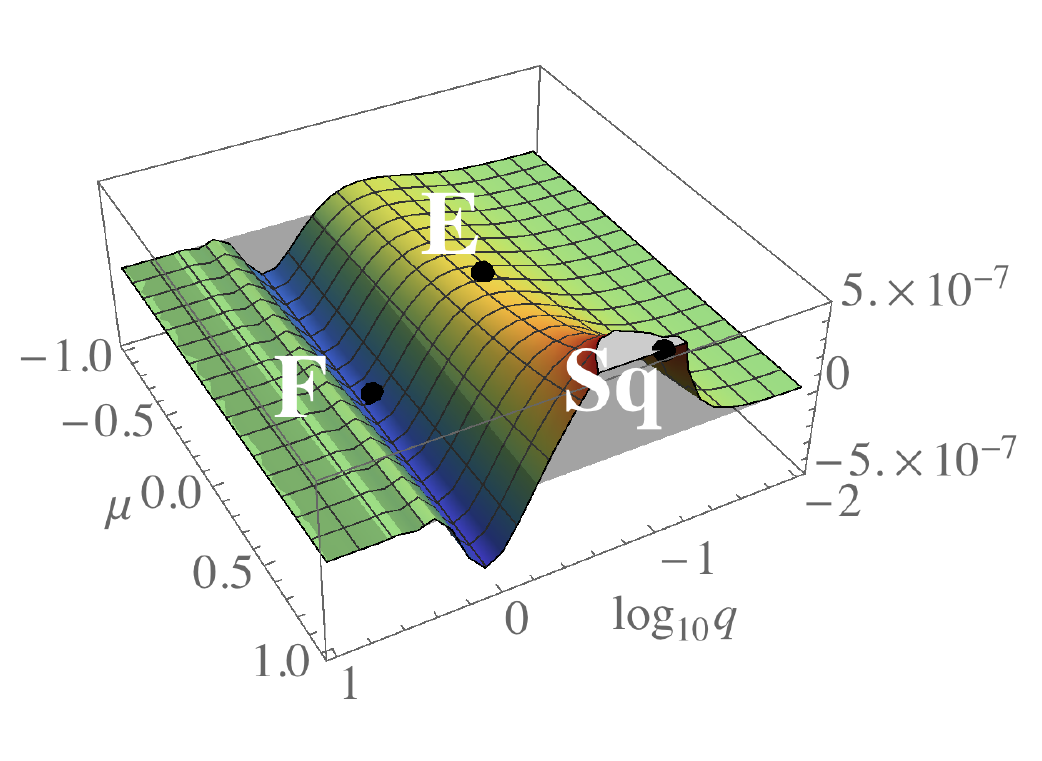}
        \label{fig:Rng_weight_m_12pt5_k_0pt1}
        \caption{$m = 10^{12.5} \, M_\odot/h$, $k = 0.1 \, h/\mathrm{Mpc}$}
    \end{subfigure}%
    \begin{subfigure}{0.5\textwidth}
    \centering
        \includegraphics[width=\textwidth]{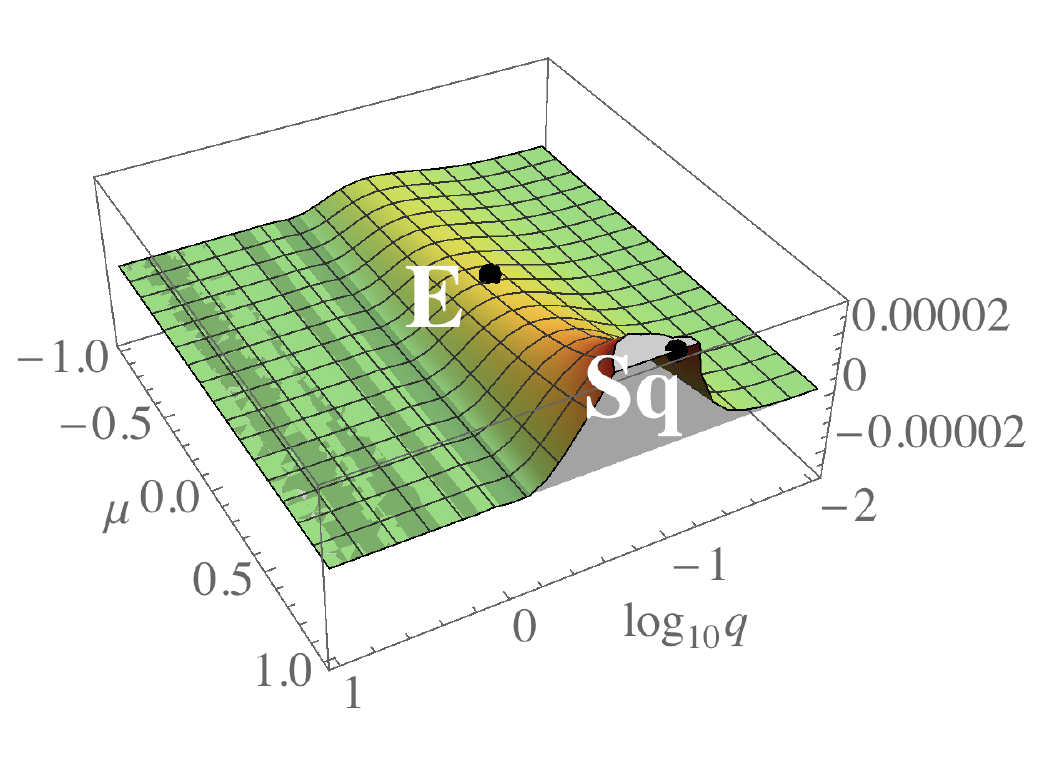}
        \label{fig:Rng_weight_m_13pt5_k_0pt1}
        \caption{$m = 10^{13.5} \, M_\odot/h$, $k = 0.1 \, h/\mathrm{Mpc}$}
    \end{subfigure}
\\    
    \begin{subfigure}{0.25\textwidth}
    \centering
        \includegraphics[width=\textwidth]{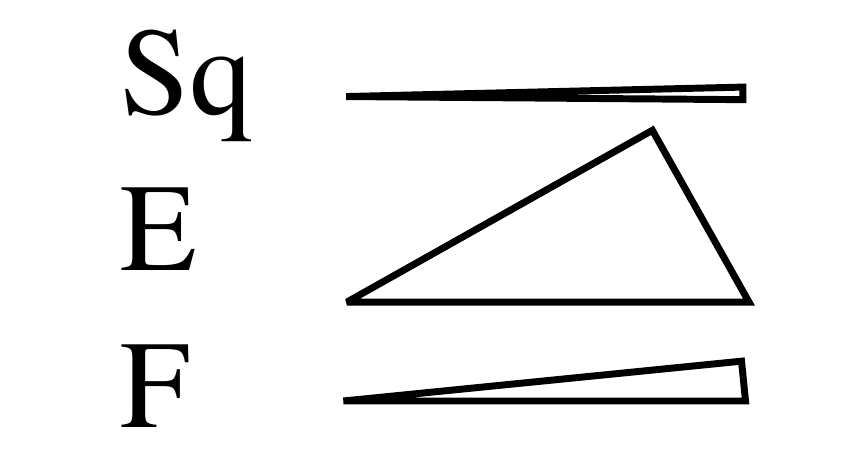}
        \label{fig:Rng_triangles}
        \caption{}
    \end{subfigure}
\\
\caption{The weight function $w_{R_{NG}}$, parameterizing the shape sensitivity in non-Gaussian corrections to the the halo mass function, as defined in \eqref{eq:Rng weight}. The function is plotted for a fixed $k = 0.1 \, h/\mathrm{Mpc}$ slice, where the weight is peaked, and two halo masses, $m = 10^{12.5} M_\odot/h$ (left) and $m = 10^{13.5} M_\odot/h$ (right).
The plots illustrate that there are non-negligible  contributions from configurations that are not significantly squeezed, like those labeled E and F.
}    
\label{fig:Rng weights}
\end{figure}

\begin{figure}[!t]
  \includegraphics[width=0.5\textwidth]{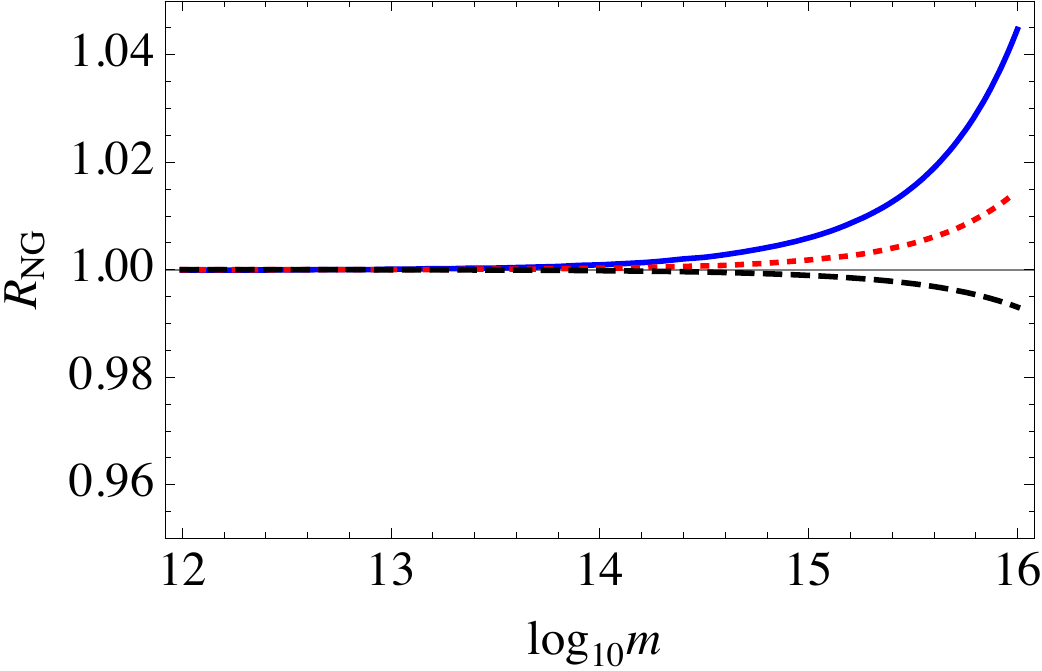}
  \includegraphics[width=0.5\textwidth]{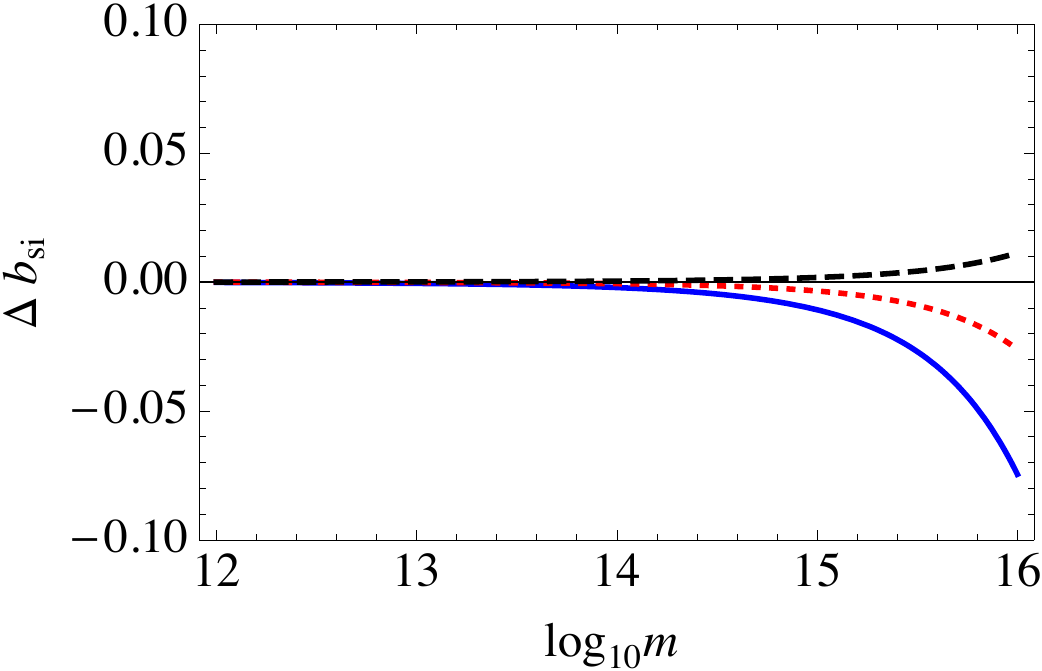}
  \caption{
The effect of the non-Gaussian corrections to the halo mass function [left panel]  and the scale-independent halo bias [right panel] on the halo power spectrum relative to the Gaussian case $P_h^{G}$, for the local [full blue], equilateral [dotted red], and orthogonal [dashed black] shapes at $z=1$ for $f_{NL}=100$.}
\label{fig:Rngbsi}
\end{figure}

To determine the shape sensitivity of the halo mass function, we can define an analogous weight function, $w_{R_{NG}}$, for $R_{NG}$:
\be
R_{NG}(m) = 1 + \int d{\ln}k \int d{\ln}q \int d\mu \; w_{R_{NG}}(k,{\ln}q,\mu,m) S(k,q,k_\mu). \label{eq:Rng weight}
\ee
We find this weight is highest and has the largest contributions to the total value of $R_{NG}$ on the $k \sim 0.1 \, h/\mathrm{Mpc}$ slice, shown in Figure \ref{fig:Rng weights}.
The figure illustrates that the weight has the largest amplitude for triangles that have $q \sim 0.1-1 \, h/\mathrm{Mpc}$.
While $w_{R_{NG}}$ does probe squeezed triangles, the weight also extends to include other configurations that are not very squeezed.
In this sense, $R_{NG}$ probes similar configurations to $\Delta b_{sd}(k)$ on mildly non-linear scales, but with a different specific triangle configuration-dependence.
Since $R_{NG}$ is an integrated measure of the shape (as it integrates over $k$ and has no remaining $k$-dependence), we might expect one can extract more information from $\Delta b_{sd}(k)$ than from $R_{NG}$. The non-Gaussian modification to the halo mass function and the scale-independent halo bias by the local, equilateral, and orthogonal templates is shown in Figure \ref{fig:Rngbsi}.  
The figure shows that both effects are more important for the most massive halos, and is typically at the few percent level.

In summary, the scale-dependent bias on large scales probes general squeezed configurations of the primordial shape, while the scale-dependent bias on small scales and the halo mass function are more sensitive to a broader range of configurations.

\begin{figure}[!t]
\centering
\includegraphics[width=0.48\textwidth]{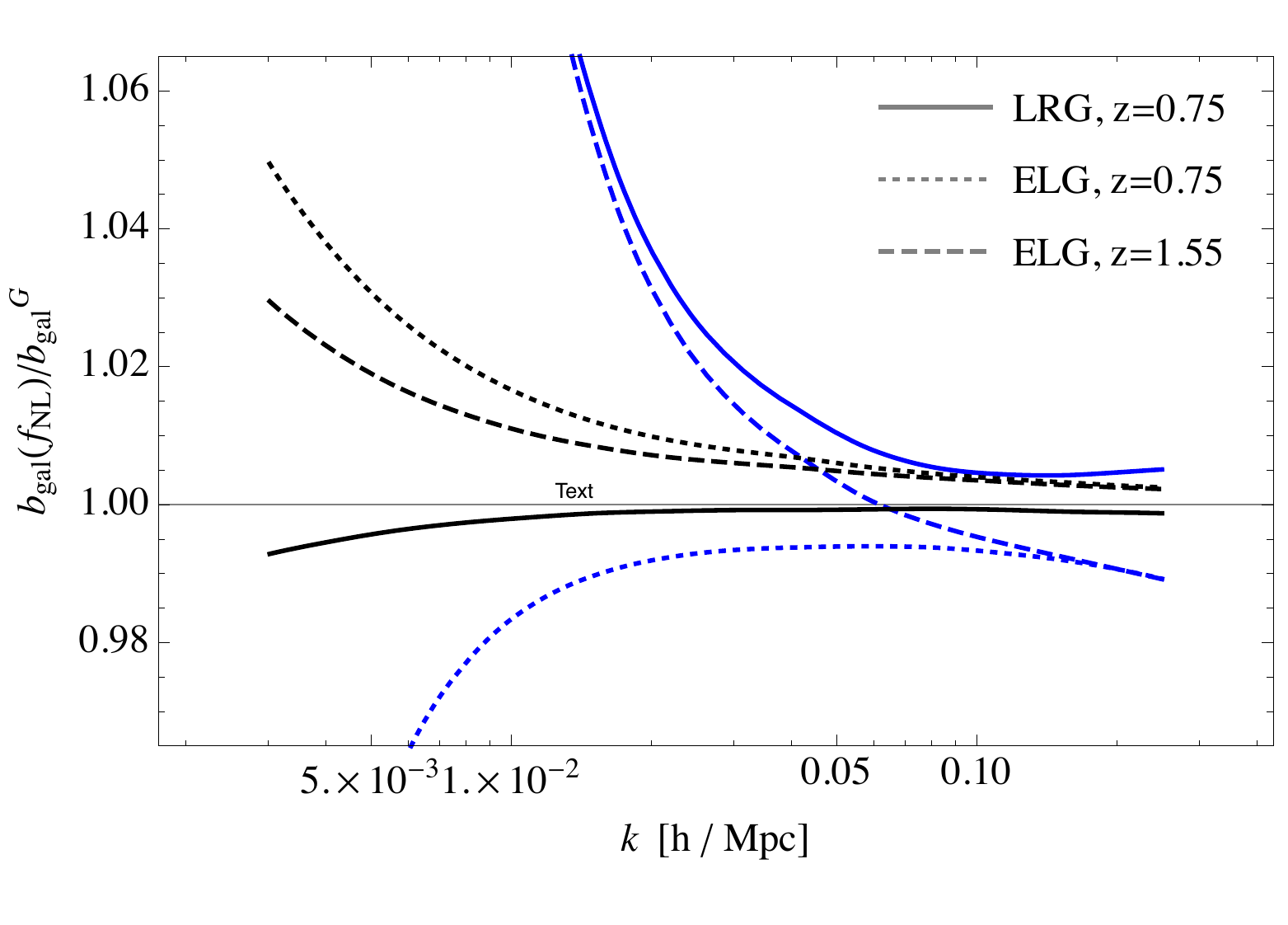}
\includegraphics[width=0.48\textwidth]{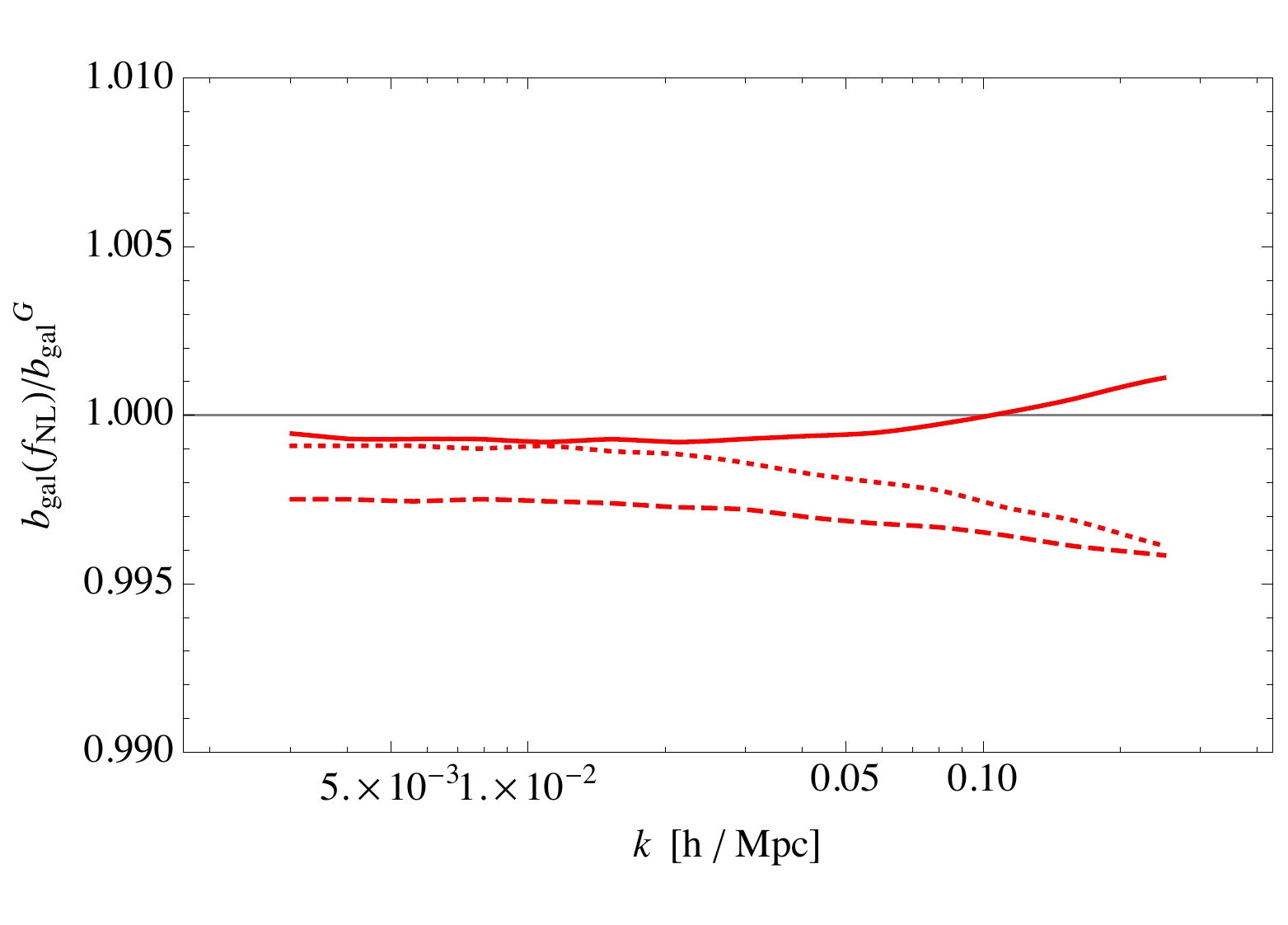}
\caption{The predicted galaxy bias as a function of scale for luminous red galaxies (LRGs) at $z=0.75$ [solid lines], and emission line galaxies (ELGs) at $z=0.75$ [dotted] and $z=1.55$ [dashed], for a variety of non-Gaussian bispectrum templates with $f_{NL}=100$: local [blue] and orthogonal [black] (left panel), as well as equilateral [red] (right panel).
}
\label{fig:bgal}
\end{figure}

Figure \ref{fig:bgal} shows that the redshift evolution and scale dependence of the galaxy bias are distinct from that of the Gaussian bias predicted using the halo model and different for both the ELG and LRG galaxy samples at the same redshift. 
These differences might be utilized to extract out information about the Gaussian component of the galaxy bias, using weak lensing \cite{Jeong:2009wi,Takeuchi:2010bc,Giannantonio:2011ya,Giannantonio:2013kqa}
and three-point statistics \cite{Sefusatti:2007ih}.
These analyses suggest 1-10\% constraints using higher order statistics, and comparable or better constraints using weak lensing might be achievable.
We consider constraints from the galaxy clustering power spectrum alone, with the possibility of additional information by imposing 1\% and 10\% constraints on the Gaussian galaxy bias, and the limiting case, of perfect Gaussian bias determination, in which the Gaussian bias is not marginalized over at all in the analysis.

\subsection{Prospective constraints from upcoming LSS and CMB surveys}
\label{subsec:fisher}

Here we discuss the results of the Fisher matrix analysis, using the approach described in Section \ref{subsec:fisher_approach}, establishing constraints on a variety of non-Gaussian templates in section \ref{subsub:specific shapes}, and the potential to distinguish between characteristics of different primordial shapes in section \ref{subsub: discriminate}.

\subsubsection{Constraints on specific templates}\label{subsub:specific shapes}

\begin{table}[!h]
\centering
\begin{tabular}{|c|c|cc|cc|cc|cc|}
\hline
\multirow{2}{*}{$\sigma(f_{NL})$}&  \multirow{2}{*}{ CMB} & \multicolumn{8}{|c|}{ Euclid-like LSS (+CMB) with $b_{gal}^G(z)$ prior}\\ 
 \cline{3-10}
&  &   \multicolumn{2}{|c|}{marg.  $b_{gal}^G$}  &  \multicolumn{2}{|c|}{10\% prior}  &  \multicolumn{2}{|c|}{1\% prior}  &  \multicolumn{2}{|c|}{known $b_{gal}^G$} 
 \\ \hline
$S^{local}$ & 3.5 & 8.4 & (3.2) & 8.4 & (3.2) & 8.3 & (3.2) & 7.1 & (3.1)
 \\ \hline
$S^{equil}$ & 43 & 250 & (42) & 230 & (42) & 64 & (36) & 21 & (19)
  \\ \hline  
$S^{orth}$ & 19 & 51 & (18) & 51 & (18) & 40 & (17) & 18 & (13)
  \\ \hline
  $S^{enf}$ & 34 & 120 & (33) & 120 & (32) & 58 & (29) & 20 & (17)
  \\ \hline
$S^{orth(2)}$, $\alpha = 8.52$ & 11 & 190 & (11) & 160 & (11) & 31 & (10) & 9.8 & (7.3)
  \\ \hline
$S^{enf(2)}$, $\alpha = 0.60$ & 100 & 1700 & (100) & 1300 & (100) & 220 & (93) & 71 & (58)
  \\ \hline
$S^{SOSF}$ & 10 & 110 & (10) & 85 & (10) & 14 & (8.4) & 4.6 & (4.2)
  \\ \hline
$S_1$ & 23 & 200 & (23) & 200 & (23) & 200 & (23) & 180 & (23)
  \\ \hline
  \multicolumn{10}{c}{}  
\\ \hline
 \multirow{2}{*}{$\sigma(f_{NL})$}&  \multirow{2}{*}{ CMB} & \multicolumn{8}{|c|}{ DESI-like LSS (+CMB) with $b_{gal}^G(z)$ prior}\\
 \cline{3-10}
&  &   \multicolumn{2}{|c|}{marg.  $b_{gal}^G$}  &  \multicolumn{2}{|c|}{10\% prior}  &  \multicolumn{2}{|c|}{1\% prior}  &  \multicolumn{2}{|c|}{known $b_{gal}^G$} 
 \\ \hline
$S^{local}$ & 3.5 & 13 & (3.4) & 13 & (3.4) & 12 & (3.4) & 7.4 & (3.2)
 \\ \hline
$S^{equil}$ & 43 & 120 & (40) & 120 & (40) & 81 & (38) & 21 & (19)
  \\ \hline  
$S^{orth}$ & 19 & 52 & (18) & 52 & (18) & 46 & (18) & 19 & (13)
  \\ \hline
  $S^{enf}$ & 34 & 100 & (32) & 100 & (32) & 74 & (31) & 20 & (17)
  \\ \hline
$S^{orth(2)}$, $\alpha = 8.52$ & 11 & 76 & (11) & 75 & (11) & 42 & (11) & 9.8 & (7.3)
  \\ \hline
$S^{enf(2)}$, $\alpha = 0.60$ & 100 & 520 & (100) & 520 & (100) & 300 & (97) & 71 & (59)
  \\ \hline
$S^{SOSF}$ & 10 & 32 & (9.8) & 31 & (9.8) & 19 & (9.1) & 4.6 & (4.2)
  \\ \hline
$S_1$ & 23 & 200 & (23) & 200 & (23) & 200 & (23) & 180 & (23)
  \\ \hline
\end{tabular}

\caption {\label{tab:benchmarks}  
Fisher constraints on $f_{NL}$ for templates with varied divergences and $\mu$-dependent properties in the squeezed limit. 
We present the forecasted constraints for a Planck-like CMB experiment, and from the halo properties of galaxies measured by a Euclid-like spectroscopic survey [top] and a DESI-like survey [bottom]. The LSS constraints are shown without CMB and with (in parentheses) and with four different assumptions on the amount of information known about the galaxy bias: assuming no knowledge and marginalizing over the bias in independent redshift bins (`marg.'), assuming a 10\% and 1\% prior knowledge, and assuming the extreme limit, in which the galaxy bias  has been measured precisely by a complementary method. }
\end{table}

Constraints for the different templates discussed in Section \ref{subsec:templates} are shown in Table \ref{tab:benchmarks} for Euclid-like and DESI-like spectroscopic surveys.
In general, templates with more strongly $k$-dependent halo biases on large scales are better measured by LSS, 
and it is clear that assumptions about the prior knowledge of the galaxy bias in each spectroscopic bin can have a major effect on the constraints.
The bias assumption is most key for the least divergent shapes: constraints degrade most when removing any assumptions on galaxy bias for the shapes with the least scale-dependent bias.
Below, we describe why this is the case, and what steps might be taken to improve constraints for the least divergent shapes.

For templates that have a scale-independent bias on large-scales, uncertainties in the Gaussian galaxy bias can easily mimic the effects of non-Gaussianity. Scale-dependent changes to the halo-model-derived $P_{gal}$  on small scales, typically smaller than $k \sim 0.1 \, h/\mathrm{Mpc}$, then provide the sole constraints. If the large scale galaxy bias can be measured then these small scale effects can give rise to $f_{NL}$ constraints that are competitive with the CMB as shown in Table  \ref{tab:benchmarks}.
One way of strengthening constraints on $f_{NL}$ is to improve our understanding of the galaxy bias, through constraints from other probes as discussed in \ref{subsec:halomodel}.
In Table \ref{tab:benchmarks} we show projected constraints on $f_{NL}$ with a range of assumptions about the Gaussian galaxy bias, including 1\% and 10\% priors roughly consistent with current constraints from the literature (e.g. \cite{Verde:2001sf,Seljak:2004sj,Gil-Marin:2014sta,DiPorto:2014yua}).

One interesting feature of the constraints in Table \ref{tab:benchmarks} is that DESI's marginalized constraints for shapes with scale-independent large-scale halo biases ($S^{equil}$, $S^{orth(2)}$, $S^{enf(2)}$, and $S^{SOSF}$) are stronger than those forecasted for Euclid.
The principal constraining power in these cases comes from DESI's LRG population, which we attribute to the LRGs having a different HOD that probes more massive halos.
At the same time, the ELG population is still better at constraining the more scale-dependent shapes like the local, orthogonal, and enfolded templates.
This seems to hint that different tracers are possibly better suited for constraints on different shapes of non-Gaussianity.

\begin{figure}[!t]
\centering
\includegraphics[width=0.475\textwidth]{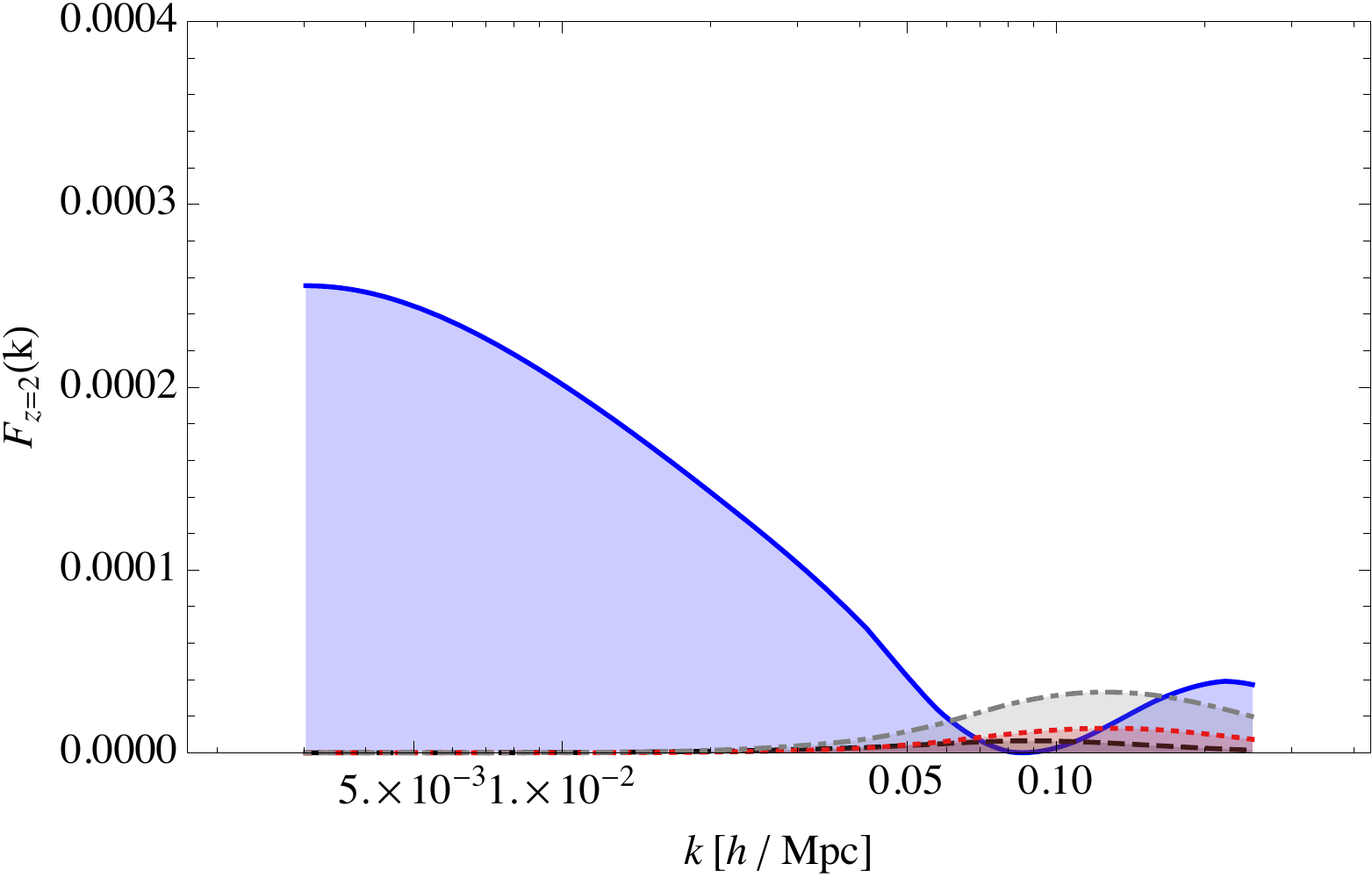}
\includegraphics[width=0.475\textwidth]{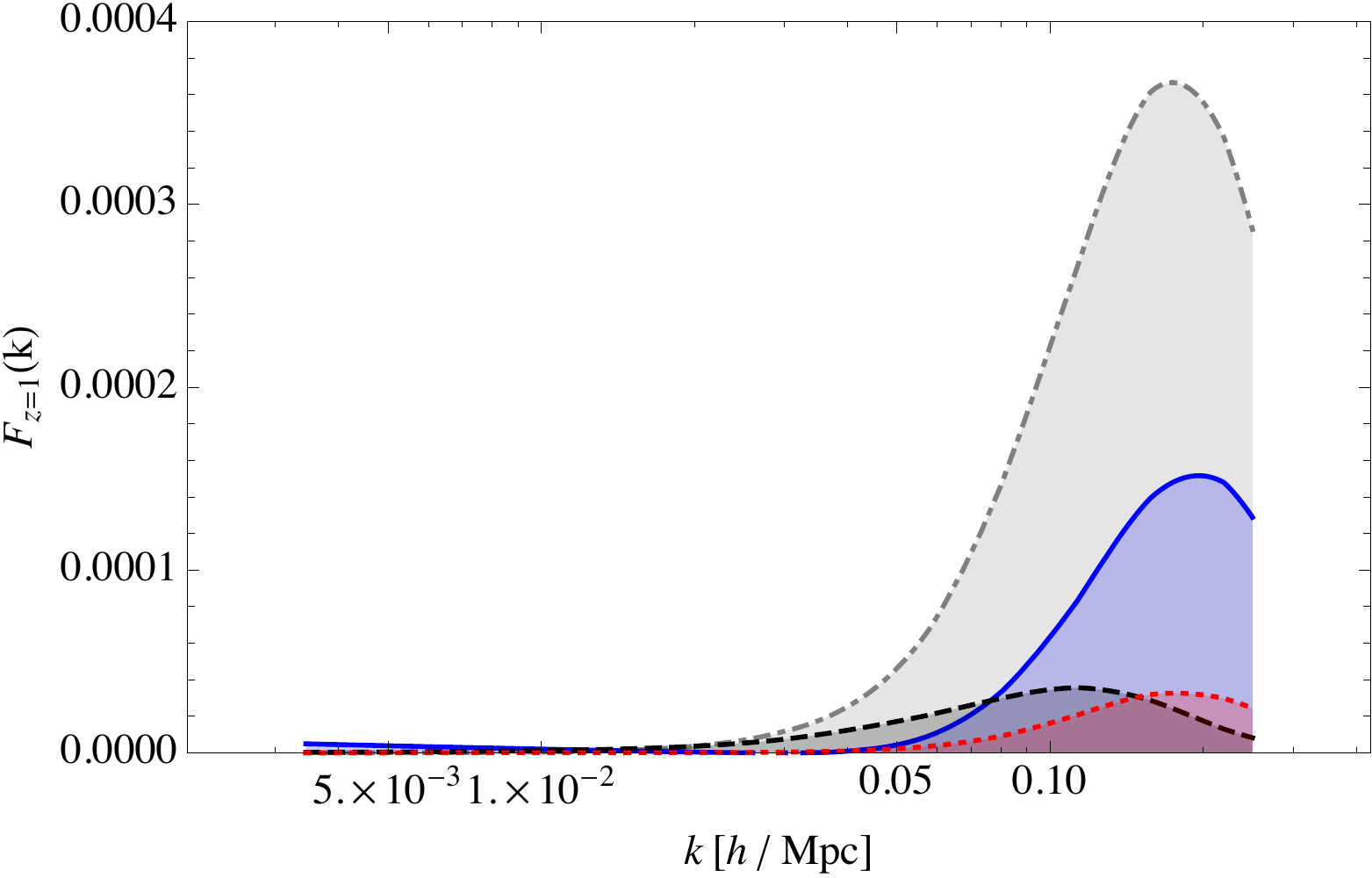}
\caption{
Contributions to a Fisher element as a function of $k$, $F_z(k)$, for $(f_{NL},f_{NL})$ in a fixed redshift slice, $z=2$ [left] or $z=1$ [right], from the local [blue], equilateral [dotted red], and orthogonal [dashed black] templates.  
Contributions to the Fisher element for $(b_{gal}^G,b_{gal}^G)$ are also shown [dot dashed gray], multiplied by a scaling factor of $10^{-8}$ in order to show on the same plot.
Unless the primordial shape has a strongly scale-dependent bias on large-scales, most of the resulting constraint on $f_{NL}$ comes from intermediate and small scales, $k \gtrsim 0.05 \, h/\mathrm{Mpc}$.
The constraint on the Gaussian galaxy bias, while it has a larger overall amplitude, its shape is qualitatively similar to the signal on small scales coming from $f_{NL}$.
}
\label{fig:Fisherk}
\end{figure}

In Figure \ref{fig:Fisherk} we show how the contribution to a Fisher element at a fixed redshift-slice varies with $k$, and it illustrates that in many cases, the principal strength of a template's constraint relies on information coming from small quasi-non-linear scales.
To illustrate the importance of small scales, we computed the LEO constraints in the case where we take more conservative assumptions about how much information we can extract from small scales.
We model this conservative case using $k_{max} = 0.15 \, h/\mathrm{Mpc}$ and $p_{NL}=1$.
In this case, the LEO marg (unmarg) constraints are: $\sigma(f_{NL}^{local}) = 9.0 \, (8.5)$, 
$\sigma(f_{NL}^{equil}) = 390 \, (33)$, 
and $\sigma(f_{NL}^{orth}) = 79 \, (23)$.
Compared to the more optimistic, default case, the marginalized fixed equilateral and orthogonal constraints weaken by factors of $\sim1.4$ and $1.5$, while the local constraint is largely unaffected.

We note that in general the constraints here are weaker than in some previous analyses of constraints from halo bias measurements, e.g. \cite{Fedeli:2010ud,Giannantonio:2011ya}.
These arise from updates lowering estimates of galaxy number counts expected from H$\alpha$ surveys \cite{Wang:2012bx,Colbert:2013ita} relative to prior expectations \cite{Geach:2009tm}, 
the inclusion of non-linear damping effects on small scales, 
and the use of the full, rather than asymptotic, forms for scale-dependence of the halo bias that alter the small scale constraints in particular for less divergent shapes.

Furthermore, the constraints also differ due to different assumptions about the Gaussian galaxy bias.
If instead of using the fiducial galaxy biases from Orsi \textit{et al.} \cite{Orsi:2009mj}, we take the galaxy bias from the halo model as our fiducial, as shown in Figure \ref{fig:gaussian_bgal_vs_z}, the resulting marginalized (unmarginalized) template constraints on LEO for a Euclid-like survey are weakened to $\sigma(f_{NL}^{local}) = 8.6 \, (7.5)$, $\sigma(f_{NL}^{equil}) = 280 \, (23)$, $\sigma(f_{NL}^{orth}) = 54 \, (20)$. 
This is consistent with previous forecasts, which also showed the sensitivity of constraints to the assumed fiducial bias \cite{Giannantonio:2011ya}.

\begin{figure}[!t]
\centering
\includegraphics[width=\textwidth]{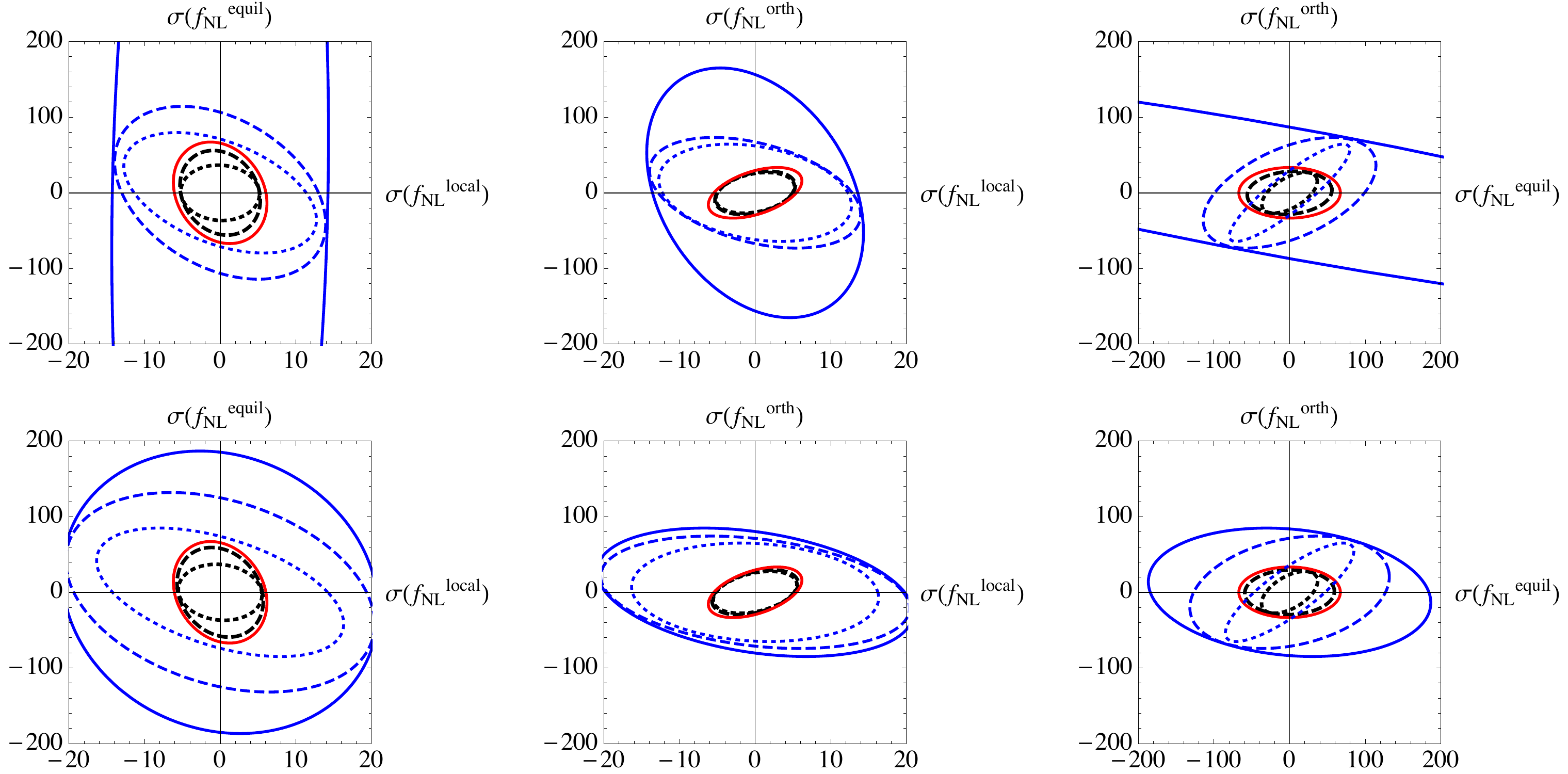}
\caption{Euclid-like [top] and DESI-like [bottom] Fisher matrix projections for the constraints on the amplitudes of contributions to a general shape constructed from local, equilateral and orthogonal (LEO) templates. 
Constraints from the halo bias only [blue], CMB only [red], and in combination [black] are shown for a range of assumptions on prior knowledge of the galaxy bias: the most conservative scenario [full lines] in which all Gaussian bias information is obtained from this dataset and $b_{gal}^G$ is marginalized over fully, 
the opposite regime [dotted lines], in which galaxy bias are wholly determined by an alternative technique, such as those discussed in section \ref{subsec:galmodel}, and is not marginalized over,
and the optimistic scenario [dashed lines] in which the Gaussian bias is well-constrained which we represent using a 1\% prior.}
\label{fig:fisher_ellipses_LEO_Euclid_Desi}
\end{figure}

\begin{figure}[!t]
\centering
\includegraphics[width=0.48\textwidth]{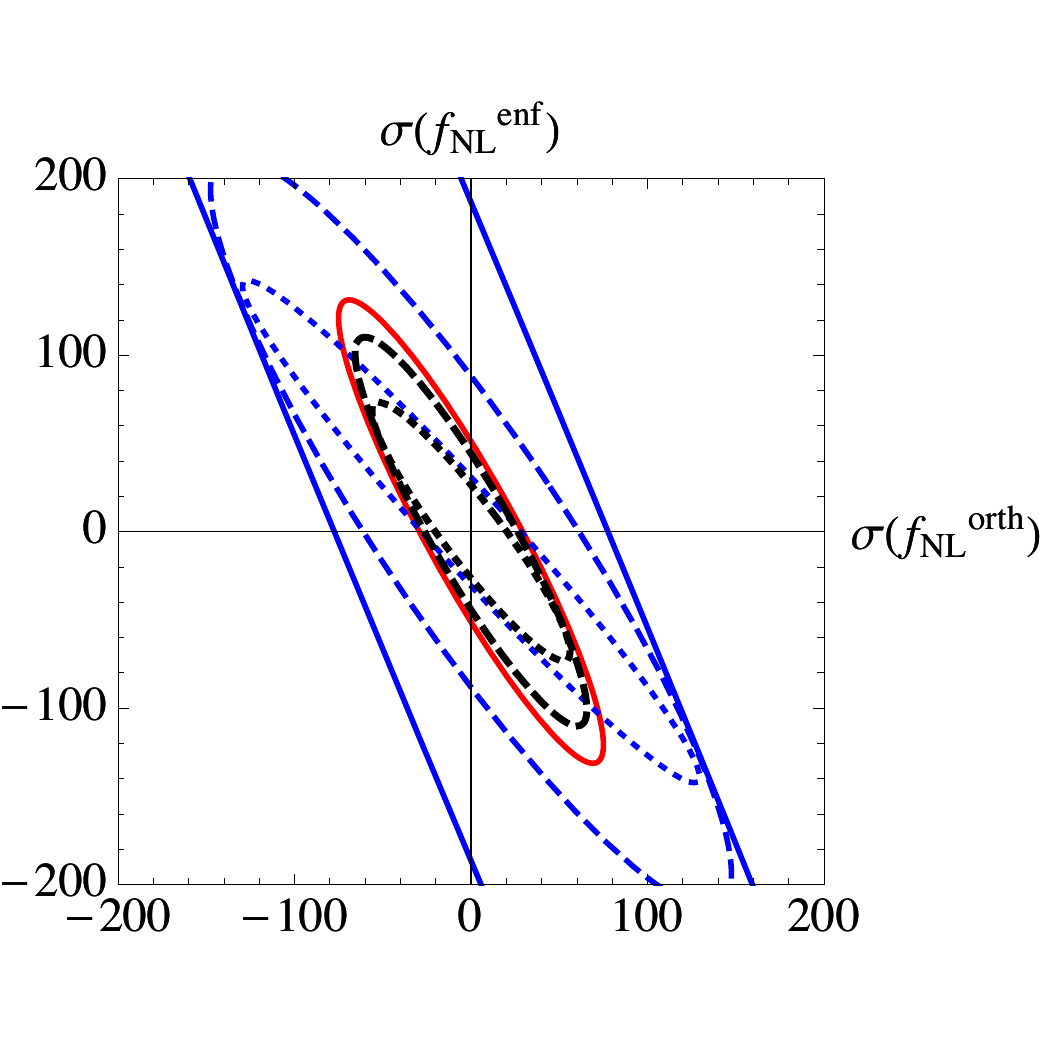}
\includegraphics[width=0.48\textwidth]{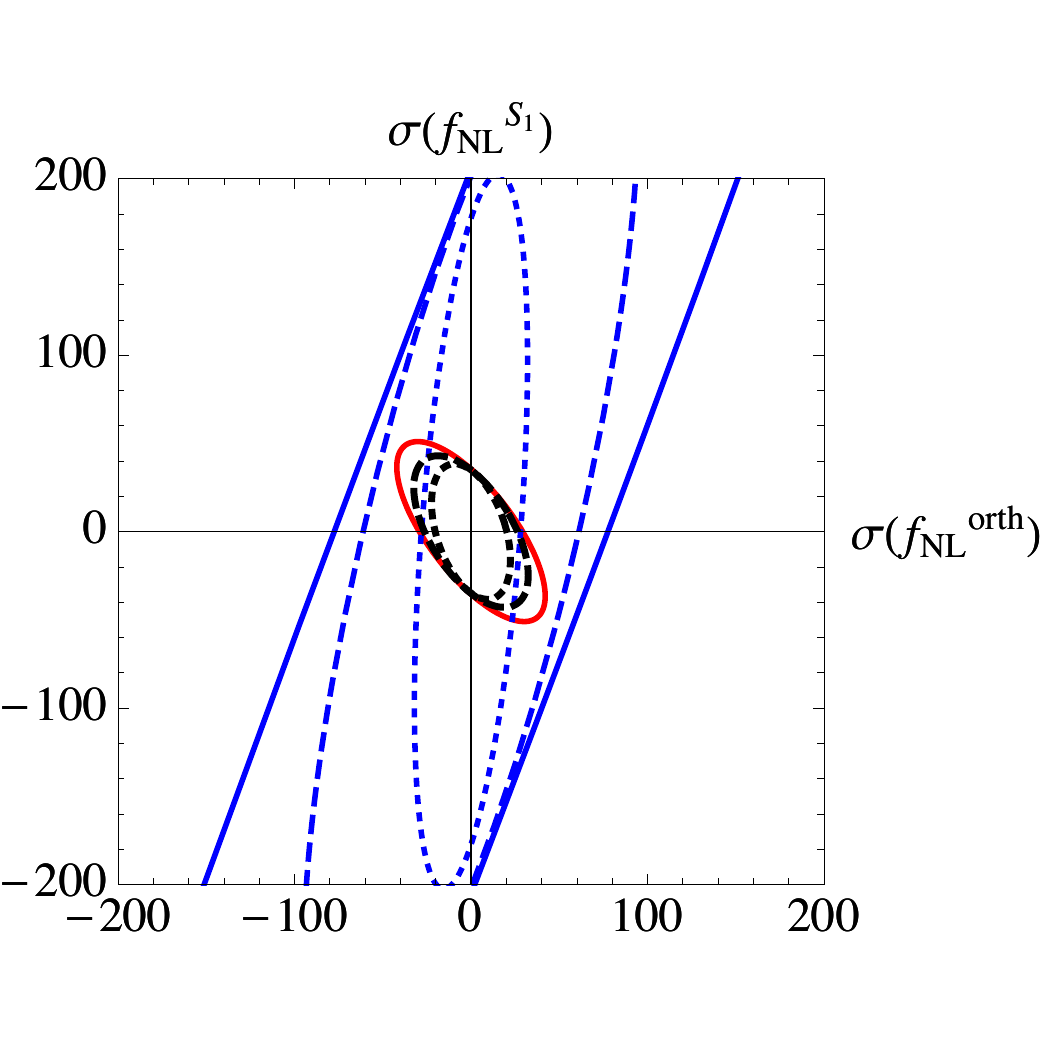}
\includegraphics[width=0.48\textwidth]{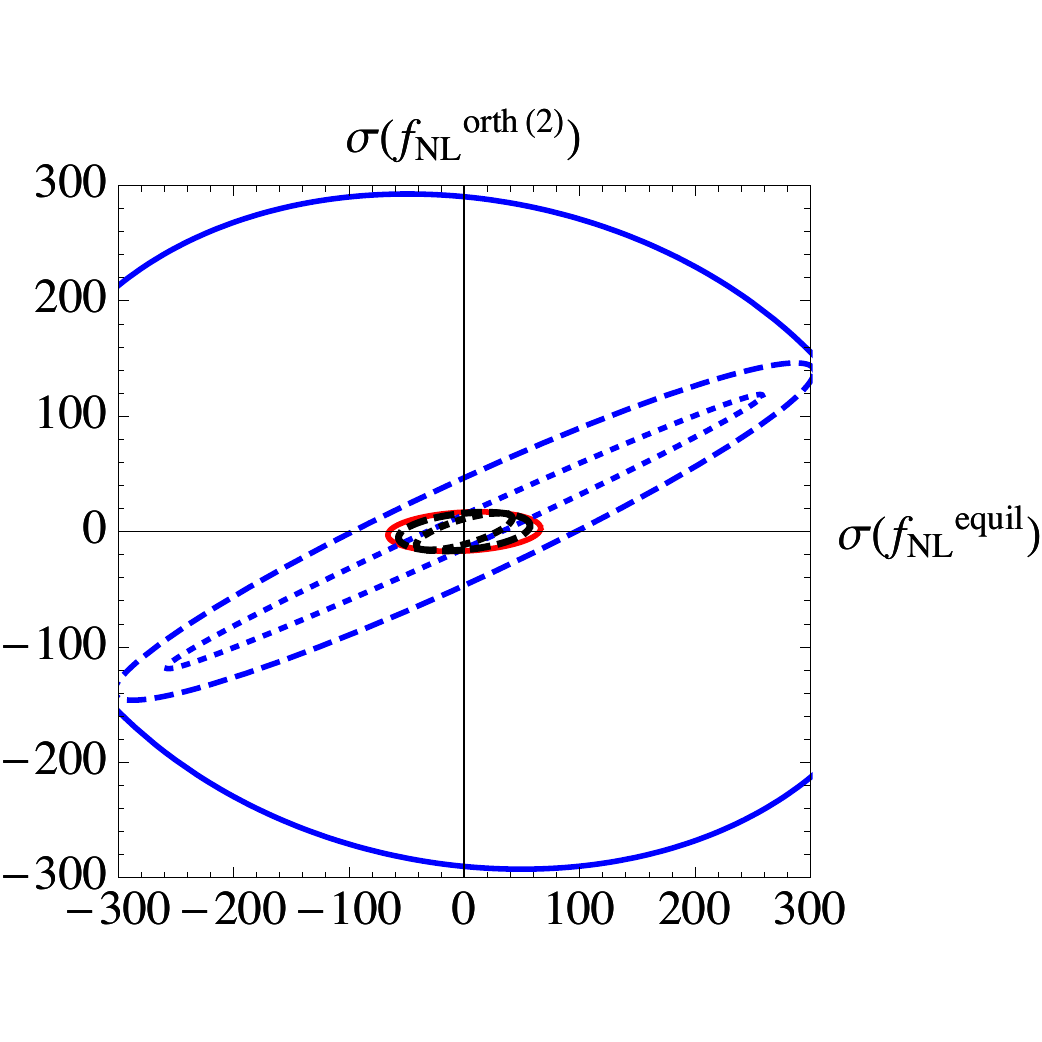}
\includegraphics[width=0.48\textwidth]{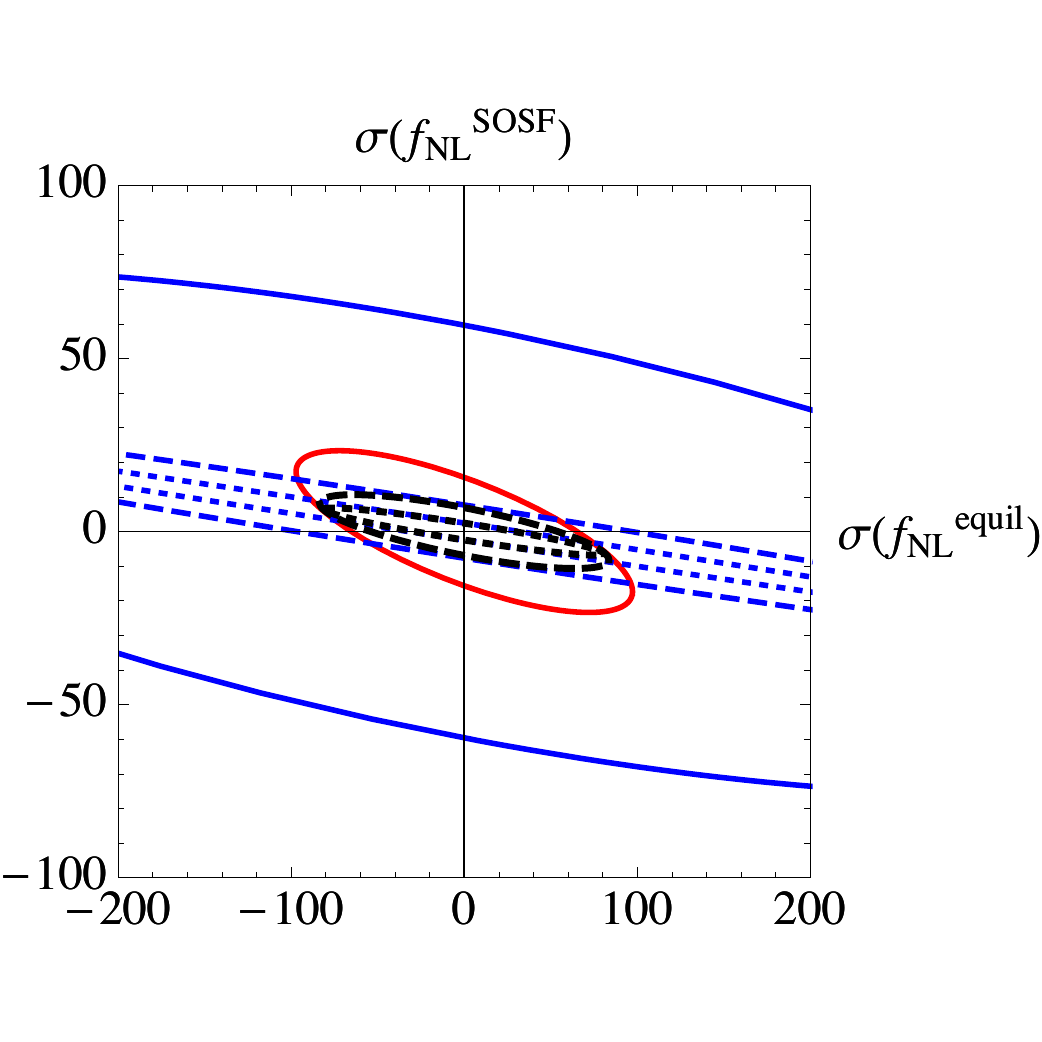}
\caption{Euclid-like Fisher matrix projections for the joint constraints on the amplitudes of a pair of shapes. Constraints from  the halo bias only [blue], CMB only [red], and in combination [black] are shown for  scenarios in which $b_{gal}^G$ is marginalized over fully [full], 
in which galaxy bias is wholly determined by an alternative technique  [dotted lines], and in which the Gaussian bias is well-constrained, which we represent using a 1\% prior [dashed lines].}
\label{fig:fisher_ellipses_beyondLEO}
\end{figure}

\subsubsection{Discriminating between shapes}\label{subsub: discriminate}

In addition to considering single templates, we are also interested in the discriminating power of the data to differentiate between templates.
We first consider the constraints on a general shape that is a linear combination of the local, equilateral, and orthogonal (LEO) templates. 

In Figure \ref{fig:fisher_ellipses_LEO_Euclid_Desi} we show the 2D marginalized constraints on these templates from the CMB, and Euclid-like and DESI-like spectroscopic galaxy surveys. With no prior information about the galaxy bias the constraints on LEO non-Gaussianity from the halo bias are projected to be much weaker than for the CMB. If galaxy bias information is known from a complementary source, such as overlapping weak lensing statistics, then the LSS gives additional constraining power in the LE and, especially, the EO planes, in which the LSS and CMB constraints are more orthogonal.
In the case of DESI, the low-$z$ LRG and high-$z$ ELG galaxies have complementary degeneracy directions in the LEO constraints, coming from the different HODs specific to red vs. blue galaxies.

We have also considered the properties of shapes that are not well described purely by the LEO basis, because of a complex divergence properties, or richer structure, examples include $S_{ortho(2)}, S_{enf(2)}, S_{1}$, and $S_{SOSF}$. 
In some cases these have similar squeezed limits, and therefore large scale halo bias properties, but are dissimilar, or have anisotropic properties, as one moves away from the squeezed limit. These provide distinctive differences at smaller scales.
In Figure \ref{fig:fisher_ellipses_beyondLEO} we consider the potential for future surveys to differentiate between pairs of such shapes.

In the case of the orthogonal and enfolded shapes, halo predictions for the two shapes are too similar to distinguish them with the galaxy power spectrum.
The degeneracy directions in the LSS and CMB data are somewhat complementary, and with knowledge of the galaxy bias, the overall LSS+CMB constraint is stronger than either data set alone.

The $S_1$ shape, has the same isosceles squeezed limit (up to a constant) as the orthogonal template but is anisotropic as it approaches the squeezed limit.
We find that the LSS constraints on a scenario looking to distinguish these two shapes are quite orthogonal to those of the CMB.
Even in the case where the Gaussian bias is marginalized over and the two shapes are highly correlated with LSS data alone ($|\rho| > 0.98$), the LSS+CMB constraint is  improved over just the CMB constraint.
With prior knowledge of the Gaussian bias, the forecasts suggest that degeneracies that exist between the anisotropic $S_1$ shape and isotropic $S_{orth}$ shape can be further broken with the inclusion of LSS data, to a correlation coefficient $|\rho| = 0.48$. 

We can also consider how well one might distinguish between models that all vanish in the squeezed limit. To do this we consider discriminating between the equilateral template and two others:  the orthogonal(2) and SOSF shapes. Because of their common large scale scale-independent bias there is a strong degeneracy, and weak constraints, for the  LSS-only constraints with no prior knowledge about the Gaussian galaxy bias. If the Gaussian bias is known well, however, the degeneracy directions between the CMB vs LSS are very complementary.

\begin{figure}[!t]
\centering
\includegraphics[width=0.45\textwidth]{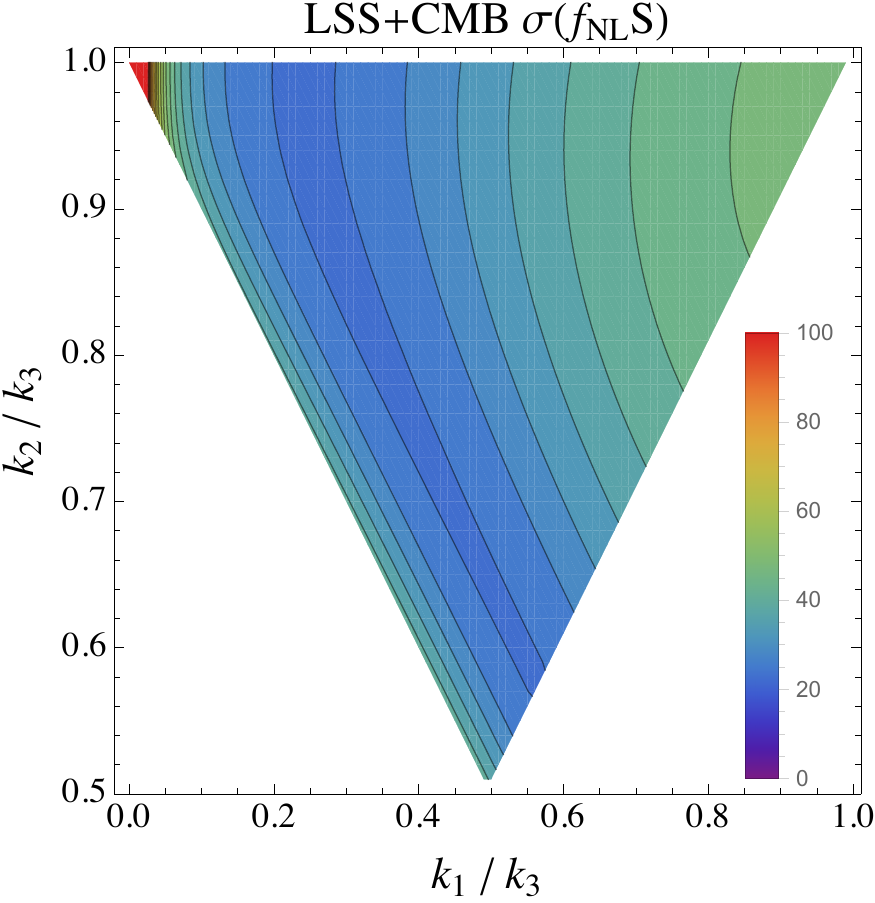}
\includegraphics[width=0.45\textwidth]{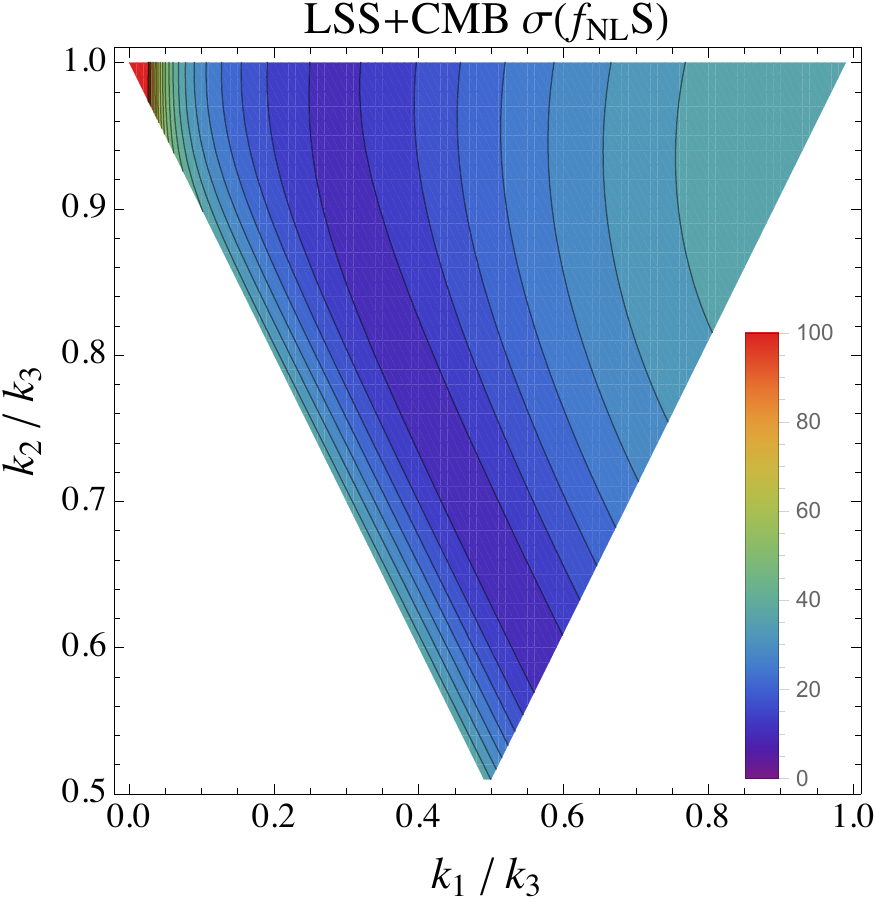}
\caption{Joint LSS+CMB $k$-dependent constraints on a general shape, assuming a basis set of local, equilateral, and orthogonal templates assuming no  [left]  and perfect [right] prior knowledge of the Gaussian bias.}
\label{fig:sigma_joint}
\end{figure}

Finally, beyond distinguishing pairs of bispectrum shapes, one could consider model-independent constraints considering a shape general constructed from a linear combination of LEO models.  As shown in Figure \ref{fig:sigma_joint}, we find $k$-dependent constraints as $\sigma(f_{NL} S(k_1,k_2,k_3)) $ are strongest in a direction that spans the local and flattened configurations.  Prior knowledge of the Gaussian bias improves the constraints but does not significantly alter the morphology of the best measured region.

%% file: sept18_halobiaspaper_discussion.tex
\section{Discussion \& Conclusions}
\label{sec:discussion}

The large number of viable inflationary theories currently in the literature warrant a careful consideration of how well upcoming galaxy surveys will be able to observationally discriminate between contending models. In this work, we focused on forecasting the ability of forthcoming galaxy power spectrum data from the spectroscopic portions of future surveys, such as Euclid and DESI, to constrain amplitudes of a variety of non-Gaussian shapes, as well as to distinguish between shapes, and put $(k_1,k_2,k_3)$-dependent constraints on a general primordial shape. In particular, the galaxy power spectrum is affected by changes to the halo bias and halo mass function that arise from non-zero primordial shapes.

Although the halo bias on large-scales generally reflects the primordial shape's behavior in the squeezed limit (i.e. if the squeezed limit of the isotropic primordial shape scales as $\propto 1/k^\alpha$, then the halo bias generally goes as $\propto 1/k^{\alpha+1}$), we find this simple correlation between late-time observables and the primordial shape is less straightforward when shapes which are anisotropic in the squeezed limit, such as $S_1$, are considered.

We  consider the effects of non-Gaussianity beyond the large scale, scale dependent halo bias, including the scale independent corrections and the effect on the  halo mass function. By defining two weighting functions,
we found that while the halo bias on large scales is a weighted average of the shape that is strongly peaked on very squeezed configurations (as expected from analytic arguments in Section \ref{subsec:squeezed}),  the halo mass function and halo bias on small scales probes more general configurations that are not very squeezed.

A significant hurdle is the degeneracy between the signature induced by non-Gaussianity in the scale-independent halo bias and the Gaussian galaxy bias.
The halo model bridges the gap between halos and their resident galaxies and, 
to a large extent, the galaxy power spectrum preserves the effect of non-Gaussianity on the halos: the large-scale behavior of $P_{gal}$ depends largely on the scale-dependent halo bias, while on smaller scales the $k$-dependence of $P_{gal}$ is a combination of all effects, from the halo bias, halo mass function and the halo occupation distribution. In theory the use of different galaxy samples, with different HODs, and redshift distributions as tracers, which map the halo information  to the galaxy bias in different ways, could allow the Gaussian and scale-independent non-Gaussian contributions to be disentangled.

Understanding the galaxy power spectrum on small quasi-non-linear scales and measuring it precisely can enrich our understanding of the primordial shape: rather than using the survey data to only constrain the squeezed limit of the primordial shape, we can gain information about other regions of the shape as well.
We include the potential to differentiate between these two effects, for example through the additional information from weak gravitational lensing, in the Fisher analysis through considering constraints in the presence of a prior on the Gaussian galaxy bias.
The forecasted constraints modeled on Euclid and DESI for a range of templates, show that for shapes with a significant scale-dependent bias, assumptions about the scale-independent bias are not critical. 
In the cases where there is not significant scale dependence, 
DESI's LRGs are a more powerful probe of $f_{NL}$ in the absence of any information about the Gaussian galaxy bias, while
knowledge of the Gaussian galaxy bias to a percent or better makes Euclid and DESI's constraints comparable, and both LSS data sets competitive with CMB constraints, such that the combined LSS+CMB constraint is better than either alone.

In addition to constraints for fixed shapes we show the forecasts for each survey's ability to distinguish between the local, equilateral, and orthogonal shapes, using LSS-data only, CMB-data only, and the combination of LSS+CMB data.
In this case, when no additional information about the Gaussian bias is included, Euclid- and DESI-like constraints are similarly weak in comparison to CMB constraints, though the two LSS surveys have different degeneracy directions in the LEO planes, due to their differing galaxy populations.
However, if the galaxy bias is constrained the Euclid-like and DESI-like surveys are comparable, with Euclid's constraints being slightly stronger. 
Additionally we find that, with strong priors on the galaxy bias, LSS and CMB constraints have complementary degeneracy directions, particularly in the $(f_{NL}^{local},f_{NL}^{equil})$ and $(f_{NL}^{equil},f_{NL}^{orth})$ planes.

We explored how well LSS data might distinguish between shapes which have similar scale-dependent biases on large scales, leveraging their quite different forms away from the squeezed limit, or their distinct, anisotropic vs isotropic, squeezed limits.
We find that with strong priors on the galaxy bias, LSS and CMB constraints have complementary degeneracy directions and the combination of the data improves upon either data set alone, but the LSS data alone with no other information about the Gaussian galaxy bias is typically not going to noticeably improve on the CMB-only constraints.

Finally, we computed the $k$-configuration-dependent constraints on general shapes via a principal component analysis.
The principal components illustrate what features of the primordial shape the galaxy power spectrum are best and worst measured, and the $k$-dependent constraints show how well we can measure different configurations of a general shape.
Without any additional knowledge about the galaxy bias, the LSS constraints do not improve on the CMB constraints.
With percent or better priors on the galaxy bias, the LSS+CMB constraints are improved accordingly, though the overall morphology of the $k$-dependent constraints on the general shape are largely unchanged. 

In conclusion, we have quantitatively demonstrated that the galaxy power spectrum can provide a valuable complementary probe of non-Gaussianity to the CMB, and can probe bispectrum configurations beyond the asymptotic squeezed limit. A key requirement to realize the full utility of the galaxy cluster measurements is knowledge of the galaxy bias. This is a rich area we plan to investigate in future work: to quantify the ability for upcoming weak lensing measurements and higher order galaxy clustering statistics to constrain the galaxy bias, as well as the potential for utilizing other tracers of the halo distribution to provide additional constraints.